\documentclass[aps,prl,reprint,superscriptaddress,longbibliography]{revtex4-2}

\usepackage{amsmath,amssymb,bm}
\usepackage{graphicx}
\usepackage[colorlinks=true,linkcolor=blue,citecolor=blue,urlcolor=blue]{hyperref}
\usepackage{xcolor}
\usepackage{float}

\usepackage{bibunits}
\providecommand{\vect}[1]{\bm{#1}}
\providecommand{\matr}[1]{\mathbf{#1}}
\providecommand{\mgreek}[1]{\bm{#1}}
\defaultbibliographystyle{apsrev4-2}
\defaultbibliography{refs}
\makeatletter
\newcommand{\clearrevtexfrontmatternotes}{\global\let\@FMN@list\@empty}
\makeatother

\begin{document}

\title{Ferron Hall effect: Transverse accumulation of polarization driven by thermal gradients in ferroelectrics}

\author{Daniel A. Bustamante Lopez}
\email{dabl@bu.edu}
\affiliation{Department of Physics, Boston University, Boston, Massachusetts 02215, USA}
\affiliation{Department of Applied Physics and Science Education, Eindhoven University of Technology, 5612 AP Eindhoven, Netherlands}

\author{Verena Brehm}
\affiliation{Department of Applied Physics and Science Education, Eindhoven University of Technology, 5612 AP Eindhoven, Netherlands}

\author{Dominik M. Juraschek}
\email{d.m.juraschek@tue.nl}
\affiliation{Department of Applied Physics and Science Education, Eindhoven University of Technology, 5612 AP Eindhoven, Netherlands}

\date{\today}

\begin{abstract}

The phonon Hall effect describes the generation of a transverse heat current in response to a longitudinal thermal gradient in a magnetic field. Here, we theoretically demonstrate that, when the lattice excitations deflected by the Hall effect carry electric dipole moments, their transverse motion produces an accumulation of electric polarization in ferroelectric materials. This accumulation is driven by lattice excitations that carry polarization, known as ferrons, and we therefore call the mechanism the \textit{ferron Hall effect}. Using atomistic lattice dynamics with parameters obtained from density functional theory, we illustrate the effect in the prototypical ferroelectric BaTiO$_3$. Our results identify ferrons as the electric-polarization analogues of magnons in transverse transport and provide a route toward thermal and magnetic manipulation of ferroic order.
\end{abstract}

\maketitle

\section{Introduction}

Hall effects are ubiquitous in solid-state physics and lie at the heart of many modern technologies \cite{Nagaosa2010AHE,Sinova2015,vonKlitzing2020,Fukami2025Orbitronics}. Conventionally, anomalous, spin, orbital, and quantum Hall effects describe the generation of a transverse charge, spin, or orbital current from a longitudinal electronic current induced by an electric field, but the modern Hall family more broadly also encompasses bosonic excitations. Examples include the Hall effect of light in optical media \cite{OnodaMurakamiNagaosa2004, KavokinMalpuechGlazov2005, LeyderRomanelliKarrEtAl2007, HostenKwiat2008, Bliokh2015}, as well as the magnon Hall effect in magnetic materials \cite{Onose2010MagnonHall, Ideue2012MagnonHall, Mochizuki2014MagnonHall, Mook2014MagnonHall,Matsumoto2014MagnonHall}. Likewise, excitations of the crystal lattice exhibit Hall responses, manifesting in the phonon Hall effect \cite{Strohm2005,Sheng2006,Inyushkin2007,Kagan2008,Zhang2010PHE,Qin2012,Barkeshli2012,Mori2014PHE,Grissonnanche2019,Saito2019PHE,Grissonnanche2020,Boulanger2020,Li2020SrTiO3PHE,Chen2020,Guo2021,Lefrancois2022,Im2022NonlinearPhononHall,Chen2022,Flebus2022,Flebus2023,Li2023BlackPhosphorusPHE,Oh2024,Chen2024,Jin2025} and phonon angular momentum Hall effect \cite{ParkYang2020,Lopez2026}, which appear to occur universally in crystalline solids. In these effects, a thermal gradient drives lattice excitations longitudinally and produces transverse transport of heat or phonon angular momentum.

\begin{figure}[b]
    \centering
    \includegraphics[width=0.85\columnwidth]{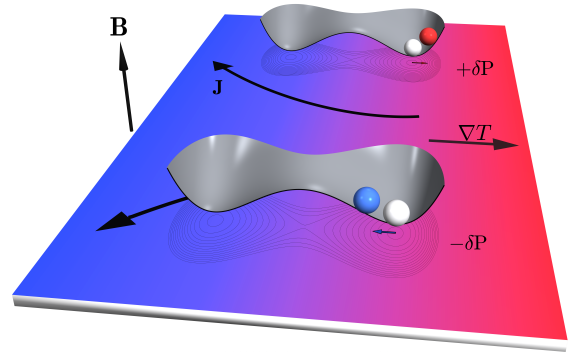}

\caption{\textbf{Ferron Hall effect.} A longitudinal thermal gradient, \(\nabla T\), drives a current, \(\mathbf{J}\), of lattice excitations carrying electric polarization, also known as ferrons, in a ferroelectric material. When time-reversal symmetry is broken by an applied magnetic field, \(\mathbf{B}\), this ferron current is deflected transversely, producing opposite polarization shifts, \(\pm\delta P\), at opposite sample edges.}
\label{fig:ferron_intro}
\end{figure}

\begin{figure}[t]
    \centering
    \includegraphics[width=\columnwidth]{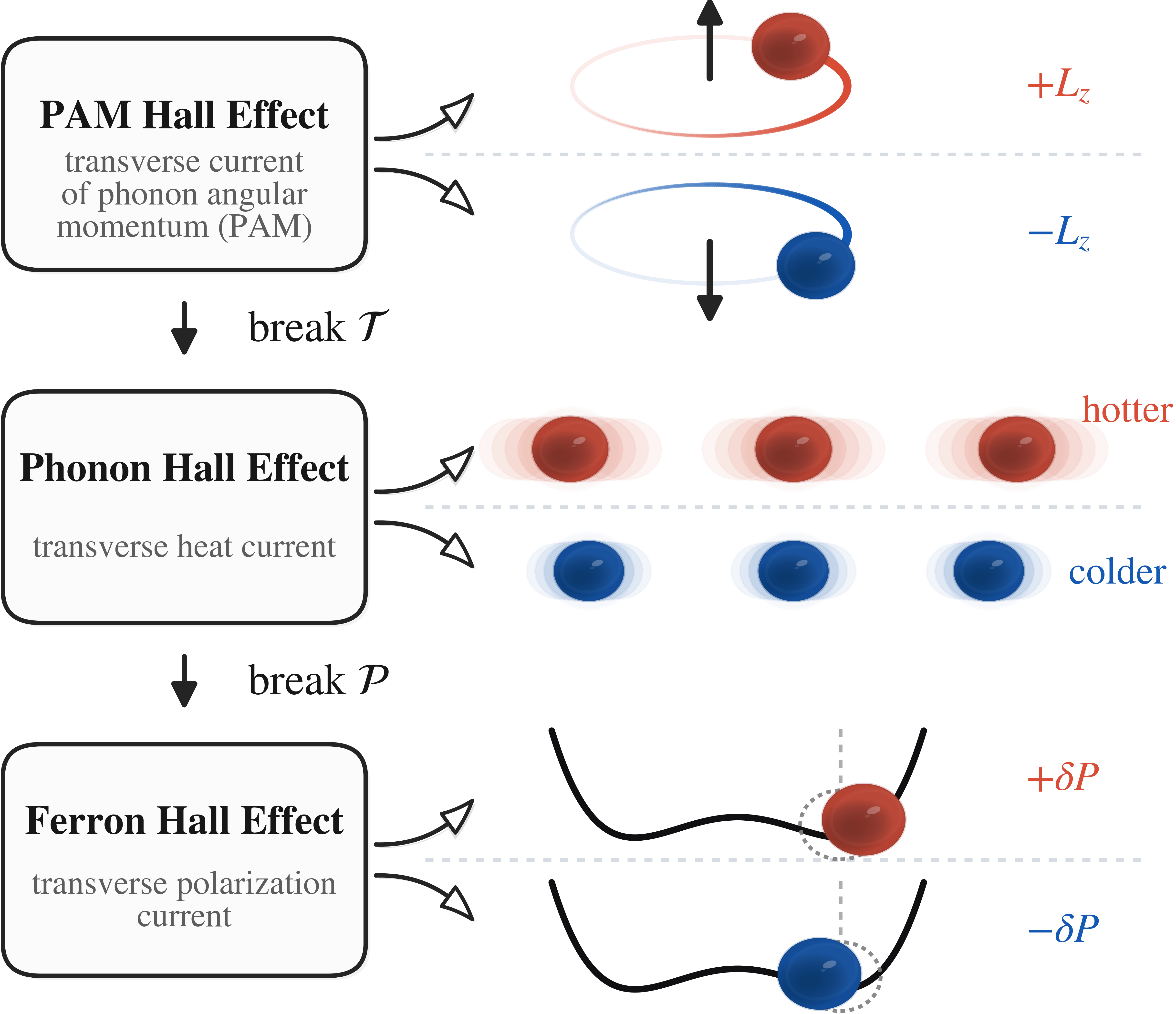}
    \caption{
    \textbf{Lattice Hall effects.}
    A longitudinal thermal gradient generates transverse responses of different lattice degrees of freedom.
    Top: Phonon angular momentum (PAM), \(\pm L_z\), accumulates with opposite signs at opposite sample edges in the form of local circular atomic motion, and has been proposed to occur universally in crystalline solids \cite{ParkYang2020,Lopez2026}.
    Middle: When time-reversal symmetry, \(\mathcal{T}\), is broken by magnetic order or an applied magnetic field, the same Hall geometry produces a transverse heat current, corresponding to the phonon Hall effect \cite{Jin2025}.
    Bottom: In a ferroelectric, broken inversion symmetry, \(\mathcal{P}\), allows ferrons to produce transverse shifts of the  polarization, \(\pm\delta P\), at opposite sample edges.
    }
    \label{fig:hall_hierarchy}
\end{figure}

Here, we complement the family of Hall effects with a previously overlooked degree of freedom: electric polarization. We demonstrate that a longitudinal thermal gradient in a ferroelectric material, together with an external magnetic field, generates a transverse polarization response, as illustrated in Fig.~\ref{fig:ferron_intro}. We formulate the effect using a stochastic lattice model governed by Langevin dynamics, with input parameters calculated from density functional theory. Within this microscopic nonequilibrium description, a longitudinal temperature gradient drives the polar modes of the ferroelectric material, while the applied magnetic field gives their motion a transverse component that changes sign when the field is reversed. In a finite sample, this transverse deflection produces an accumulation of polarization at the sample edges. This completes a hierarchy of lattice Hall effects, in which the transverse channel is heat, phonon angular momentum, or electric polarization, as summarized in Fig.~\ref{fig:hall_hierarchy}. The accumulation is driven by lattice excitations carrying electric dipole moments, also known as ferrons \cite{tangExcitationsFerroelectricOrder2022,zhouSurfaceFerronExcitations2023,bauerPolarizationTransportFerroelectrics2023, tangElectricAnalogMagnons2024, shenObservationFerronTransport2025, zhaoRoleFerronsHeat2025,morozovskaFlexocouplinginducedPhononsFerrons2025,Choe2026Ferrons,Jana2026Ferrons,Tang2026,Pols2026,Subedi2026Ferrons}, and therefore represents a \textit{ferron Hall effect}. We demonstrate the effect for the prototypical ferroelectric BaTiO$_3$.

Our study is motivated by recent work on ferron-induced polarization transport in ferroelectrics \cite{Bauer2021FerroelectricTransport,bauerPolarizationTransportFerroelectrics2023,tangElectricAnalogMagnons2024,shenObservationFerronTransport2025,zhaoRoleFerronsHeat2025}, which has led to the emerging field of \textit{ferronics}, in analogy to magnonics, which deals with magnetization transport in magnetic insulators \cite{bauerMagnonicsVsFerronics2022}.

\section{Theoretical formalism}

We begin by developing the stochastic lattice model used to describe the Hall response of electric polarization. The low-energy lattice dynamics are described in terms of the polar soft mode of the ferroelectric. This mode consists of ionic displacements that connect the centrosymmetric paraelectric structure to the inversion-symmetry-broken ferroelectric state and therefore generate the ferroelectric polarization. In the ferroelectric state, this mode is intrinsically anharmonic and carries a net electric dipole moment, also known as ferron \cite{Tang2022Excitations, bauerPerspectiveFerrons2023}.  For unit cell \(i\), we represent the local amplitude and direction of this polar distortion by the local-mode coordinate \(\mathbf{u}_i=(u_{i,x},u_{i,y},u_{i,z})\). The corresponding ionic displacements are obtained from the soft-mode eigenvectors, which also define the projected mode effective charge, mass, and angular-momentum tensors introduced below.

For a given ferroelectric state with polarization direction \(\hat{\mathbf{e}}_{\parallel}\), we decompose the local mode as
\(\mathbf{u}_i=u_{i,\parallel}\hat{\mathbf{e}}_{\parallel}+\mathbf{u}_{i,\perp}\), where
\(u_{i,\parallel}=\hat{\mathbf{e}}_{\parallel}\cdot\mathbf{u}_i\). The local double-well minima are located at
\(\mathbf{u}_i=\pm u_0\hat{\mathbf{e}}_{\parallel}\), with \(u_0\) representing the equilibrium soft-mode amplitude. Thus \(u_{i,\parallel}\) controls the local amplitude of polarization, while \(\mathbf{u}_{i,\perp}\) describes fluctuations transverse to the chosen ferroelectric polarization direction.

With this local-mode coordinate, the local polarization, kinetic energy, and phonon angular momentum are
\begin{equation}
\mathbf{P}_i
=
\frac{\mathbf{Z}\mathbf{u}_i}{\Omega}, ~~
E_{{\rm kin},i}
=
\frac{1}{2}\dot{\mathbf{u}}_i^{\rm T}\mathbf{M}\dot{\mathbf{u}}_i,
~~
L_{i,k}
=
\mathbf{u}_i^{\rm T}
\boldsymbol{\Lambda}^{(k)}
\dot{\mathbf{u}}_i .
\label{eq:observables}
\end{equation}
Here \(\Omega\) is the volume of the primitive unit cell, \(\mathbf{Z}\) is the mode effective charge tensor, \(\mathbf{M}\) is the projected mode-mass matrix, and \(\boldsymbol{\Lambda}^{(k)}\) is the projected angular-momentum matrix for Cartesian component \(k\in\{x,y,z\}\). These  tensors are constructed from the soft-mode eigenvectors, Born effective charges, and atomic masses, as described in Supplemental Material \cite{SUPP}.

The energy-conserving part of the local-mode dynamics is generated by the effective Hamiltonian
\begin{equation}
\mathcal{H}
=
E_{{\rm kin}}
+
V_{\rm on}
+
V_{\rm sr}
+
V_{\rm dip},
\label{eq:hamiltonian}
\end{equation}
where \(E_{{\rm kin}}=\sum_i E_{{\rm kin},i}\) is the total kinetic energy of the local modes. The on-site potential \(V_{\rm on}\) describes the local ferroelectric instability of each unit cell, \(V_{\rm sr}\) describes short-range interactions between local-mode distortions in different unit cells, and \(V_{\rm dip}\) describes the long-range electrostatic interaction generated by the local dipole moment, \(\mathbf{p}_i=\mathbf{Z}\mathbf{u}_i\). The on-site potential is written as
\begin{equation}
V_{\rm on}
=
\sum_i
\left[
\frac{A_2}{2}u_{i,\parallel}^2
+
\frac{B_4}{4}u_{i,\parallel}^4
+
\frac{C_6}{6}u_{i,\parallel}^6
+
\frac{\kappa_\perp}{2}
|\mathbf{u}_{i,\perp}|^2
\right],
\label{eq:onsite_potential}
\end{equation}
where \(A_2\), \(B_4\), and \(C_6\) parameterize the local double-well potential along the ferroelectric coordinate, while \(\kappa_\perp\) sets the restoring curvature for transverse fluctuations. This local-mode form follows the first-principles effective-Hamiltonian description of perovskite ferroelectrics, in which a low-energy polar coordinate describes the ferroelectric instability \cite{Zhong1994,Zhong1995PRB_BaTiO3,Ghosez1998}. Detailed definitions of \(V_{\rm sr}\) and \(V_{\rm dip}\), together with numerical parameters, are given in Supplemental Material \cite{SUPP}.

The forces entering the equations of motion are obtained from the potential-energy terms as \(\mathbf{F}_i=-\partial V/\partial \mathbf{u}_i\). To describe nonequilibrium transport, we add the magnetic-field-induced gyroscopic force \(\mathbf{G}_{\Omega_B}\dot{\mathbf{u}}_i\), Langevin damping with coefficient \(\gamma_i\), and the stochastic thermal force \(\boldsymbol{\eta}_i(t)\), following related stochastic treatments of nonequilibrium lattice dynamics \cite{Juraschek2020_1,Caprini2024,Lopez2026}. The stochastic equation of motion is then
\begin{equation}
\mathbf{M}\ddot{\mathbf{u}}_i
=
\mathbf{F}^{\rm on}_i
+
\mathbf{F}^{\rm sr}_i
+
\mathbf{F}^{\rm dip}_i
+
\mathbf{G}_{\Omega_B}\dot{\mathbf{u}}_i
-
\gamma_i\mathbf{M}\dot{\mathbf{u}}_i
+
\boldsymbol{\eta}_i(t).
\label{eq:eom}
\end{equation}

The on-site contribution is
\begin{equation}
\mathbf{F}^{\rm on}_i
=
-
\left(
A_2+B_4u_{i,\parallel}^2+C_6u_{i,\parallel}^4
\right)
u_{i,\parallel}\hat{\mathbf{e}}_{\parallel}
-
\kappa_\perp\mathbf{u}_{i,\perp}.
\label{eq:onsite_force}
\end{equation}
The short-range interaction between local-mode distortions in different unit cells gives the pair-difference force
\begin{equation}
\mathbf{F}^{\rm sr}_i
=
-
\sum_{\mathbf{R}}
\mathbf{K}^{\rm sr}(\mathbf{R})
\left(
\mathbf{u}_{i+\mathbf{R}}-\mathbf{u}_i
\right),
\label{eq:Fsr}
\end{equation}
where \(\mathbf{R}\) connects primitive unit cells and \(\mathbf{K}^{\rm sr}(\mathbf{R})\) is the \(3\times3\) kernel obtained by projecting first-principles interatomic force constants onto the soft-mode basis. This term captures the short-range energy cost of relative polar distortions between unit cells.

The long-range electrostatic force arises because each local mode carries a dipole moment \(\mathbf{p}_i\). Since the dipole-dipole interaction is long-ranged, it is convenient to express the force in reciprocal space as
\begin{equation}
\mathbf{F}^{\rm dip}(\mathbf{q})
=
-
\mathbf{K}^{\rm dip}(\mathbf{q})\mathbf{u}(\mathbf{q}),
\qquad
\mathbf{F}^{\rm dip}(\mathbf{q}=\mathbf{0})=0,
\label{eq:Fdip}
\end{equation}
where \(\mathbf{u}(\mathbf{q})\) is the Fourier transform of the local-mode field. For finite wave vector \(\mathbf{q}\), electronic screening is included through the high-frequency dielectric tensor \(\boldsymbol{\epsilon}_{\infty}\), yielding the screened dipolar kernel \cite{Zhong1994Zstar,Ghosez1998DynamicalCharges}
\begin{equation}
\mathbf{K}^{\rm dip}(\mathbf{q})
=
\frac{4\pi k_e}{\Omega}
\frac{
\left(\mathbf{Z}^{\rm T}\mathbf{q}\right)
\left(\mathbf{Z}^{\rm T}\mathbf{q}\right)^{\rm T}
}{
\mathbf{q}^{\rm T}
\boldsymbol{\epsilon}_{\infty}
\mathbf{q}
},
\label{eq:Kdip}
\end{equation}
where \(k_e\) is the Coulomb constant. The numerator projects the mode-induced dipole density onto the longitudinal component along \(\mathbf{q}\), while the denominator accounts for the screened Coulomb response of the electronic background. The \(\mathbf{q}=0\) component corresponds to a spatially uniform macroscopic polarization, whose electrostatic contribution depends on electrical boundary conditions. We therefore set this component to zero, corresponding to no imposed macroscopic depolarizing field, and retain the finite-\(\mathbf{q}\) depolarizing fields associated with polarization inhomogeneities. The finite-\(\mathbf{q}\) force field is then Fourier transformed back to real space.

The magnetic field is included phenomenologically through a gyroscopic force of the form generated by Zeeman-like couplings between phonon angular momentum and a magnetic field or magnetic order \cite{Bistoni2021,Geilhufe2022,Saparov2022,Juraschek2022_giantphonomag,Bonini2023,Ren2024,Chaudhary2024,Zhang2025MolecularBerryCurvature},
\begin{equation}
\mathbf{G}_{\Omega_B}
=
\Omega_B
\mathbf{M}^{1/2}
\mathbf{C}
\mathbf{M}^{1/2},
\quad
\mathbf{C}
=
\begin{pmatrix}
0 & -\hat{B}_z & \hat{B}_y\\
\hat{B}_z & 0 & -\hat{B}_x\\
-\hat{B}_y & \hat{B}_x & 0
\end{pmatrix}.
\label{eq:gyro_matrix}
\end{equation}
Here \(\Omega_B\) is an effective coupling rate and \(\mathbf{C}(\hat{\mathbf{B}})\) is an antisymmetric matrix determined by the magnetic-field direction \(\hat{\mathbf{B}}\). Reversing the magnetic field reverses the sign of \(\mathbf{C}(\hat{\mathbf{B}})\) and therefore the sign of the gyroscopic coupling. Since \(\mathbf{C}(\hat{\mathbf{B}})\) is antisymmetric, \(\mathbf{G}_{\Omega_B}\) is also antisymmetric, and the magnetic force satisfies \(\dot{\mathbf{u}}_i^{\rm T}\mathbf{G}_{\Omega_B}\dot{\mathbf{u}}_i=0\). This force therefore changes the direction of the local-mode velocity without changing its kinetic energy. It breaks time-reversal symmetry, as required for a Hall response, but does not act as a source or sink of energy.

Nonequilibrium driving is introduced through a spatially varying Langevin bath. The local temperature \(T_i\) defines the thermal profile, and the damping coefficient \(\gamma_i\) may vary spatially. The stochastic force satisfies
\begin{equation}
\left\langle
\eta_{i,\alpha}(t)\eta_{j,\beta}(t')
\right\rangle
=
2\gamma_i k_{\rm B}T_i
M_{\alpha\beta}
\delta_{ij}\delta(t-t'),
\label{eq:noise_corr}
\end{equation}
whereas $\left\langle \boldsymbol{\eta}_i(t) \right\rangle = 0$. Here \(j\) is a second unit-cell index, \(\alpha,\beta\in\{x,y,z\}\), \(k_{\rm B}\) is the Boltzmann constant, \(\delta_{ij}\) is the Kronecker delta, and \(\delta(t-t')\) is the Dirac delta function. This noise correlator is the fluctuation-dissipation relation for the local Langevin bath written in the projected mass coordinates. To solve the Langevin dynamics, we integrate Eq.~\eqref{eq:eom} with a symmetric BAOAB splitting: the forces are applied through velocity half-steps, the gyroscopic term is solved exactly as a rotation of the local-mode velocity that preserves the kinetic energy, and the Langevin step is propagated with the exact Ornstein--Uhlenbeck update~\cite{Strang1968,Gillespie1996,LeimkuhlerMatthews2013,LeimkuhlerMatthews2015}.

\section{Ferron Hall effect in $\text{BaTiO}_3$}

We now solve the stochastic equations of motion to demonstrate the ferron Hall effect in BaTiO$_3$. The simulations are performed on a three-dimensional lattice of the local-mode coordinates \(\mathbf{u}_i\) introduced above, which represent the polar soft-mode distortions in each unit cell. The numerical parameters are obtained from density functional theory and provided, together with computational details, in Supplemental Material~\cite{SUPP}. Periodic boundary conditions are imposed along the longitudinal \(x\) and the out-of-plane \(z\) directions, while the transverse \(y\) direction is kept finite to resolve edge accumulation. The temperature profile consists of a hot central band between colder outer regions, producing oppositely directed longitudinal temperature gradients on the two sides of the hot band. From the resulting nonequilibrium steady state, we compute spatial maps of the local polarization \(\mathbf{P}_i\), kinetic energy \(E_{{\rm kin},i}\), and out-of-plane phonon angular momentum \(L_{i,z}\), as defined in Eq.~\eqref{eq:observables}.

\begin{figure*}[t]
    \centering
    \includegraphics[width=0.9\textwidth]{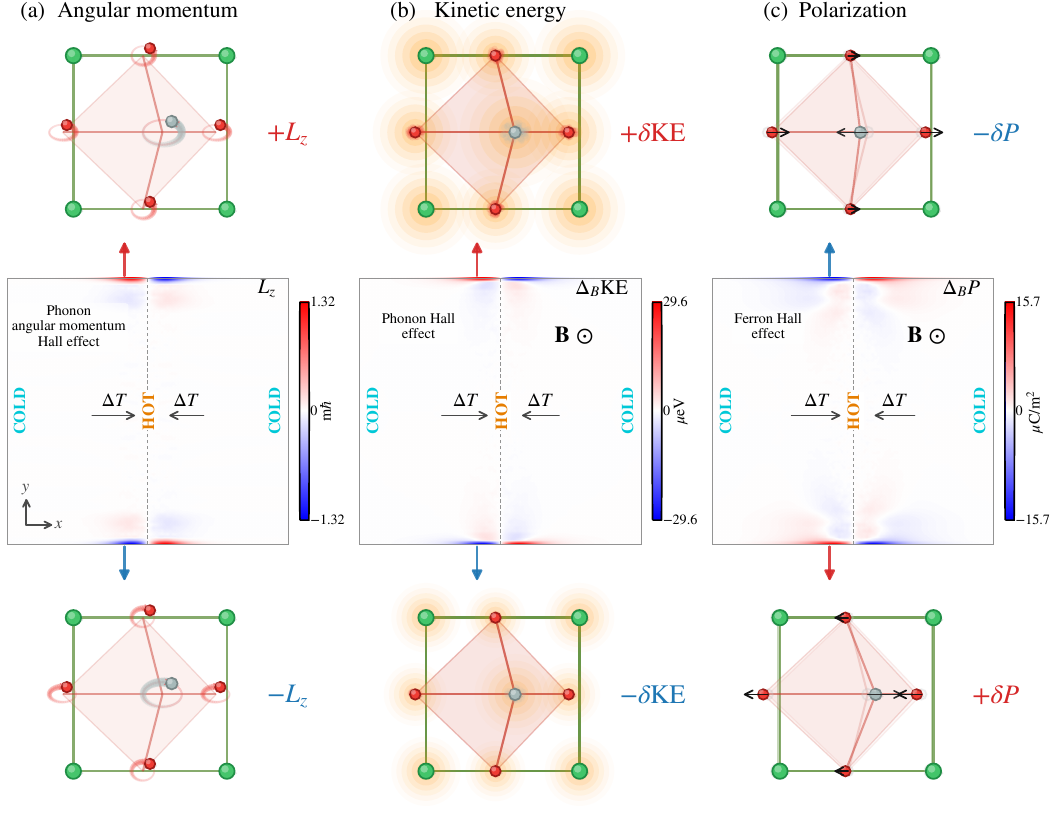}
    \caption{
    \textbf{Transverse responses of driven polar lattice dynamics in ferroelectric BaTiO$_3$.}
    \textbf{(a)} At zero magnetic field, the longitudinal thermal drive generates a transverse accumulation of out-of-plane phonon angular momentum \(L_z\), with opposite signs at opposite transverse edges and reversal on the two sides of the hot band. This is the phonon-angular-momentum Hall response.
    \textbf{(b)} For \(\mathbf{B}\parallel\hat{\mathbf z}\), the field-odd local kinetic energy \(\Delta_B\mathrm{KE}\) reveals a transverse thermal response, as expected from a phonon-Hall-type deflection of heat flow in the driven polar lattice.
    \textbf{(c)} The field-odd polarization response \(\Delta_B P\) shows opposite-sign shifts of the longitudinal ferroelectric polarization at the transverse edges. This odd-in-field transverse polarization accumulation is the ferron Hall effect. The schematics above and below the maps indicate the microscopic character and sign of each response.
    }
    \label{fig:BTO_profiles}
\end{figure*}

In Fig.~\ref{fig:BTO_profiles}, we summarize the transverse responses obtained from the stochastic lattice model for BaTiO$_3$. The dashed vertical line marks the center of the hot band, so the left and right sides of each map correspond to opposite signs of the longitudinal temperature gradient. The panels show the response of the same local-mode observables introduced in Eq.~\eqref{eq:observables}: phonon angular momentum, kinetic energy, and electric polarization.

At zero magnetic field, the thermal gradient produces a transverse angular-momentum accumulation, shown in Fig.~\ref{fig:BTO_profiles}(a). The map of \(L_z\) displays opposite signs at the two transverse edges of the sample, corresponding to local circular ionic motion with opposite handedness. The sign of the edge accumulation further reverses on opposite sides of the hot band, where the longitudinal temperature gradient changes direction. The edge accumulation reaches \(10^{-3}\hbar\) per unit cell, consistent with previous estimates \cite{ParkYang2020,Lopez2026}. This is the phonon angular momentum Hall response, which has been proposed to occur generally in crystalline solids.

We then turn on the effective gyroscopic term for \(\mathbf{B}\parallel\hat{\mathbf z}\), corresponding to a Zeeman-like coupling between phonon angular momentum and magnetic field or magnetization. The applied magnetic field breaks time-reversal symmetry, and reversing the field is represented in the simulations by reversing the sign of \(\Omega_B\). To isolate the response that is odd in the magnetic field, we use \(\Delta_B X=[X(+\Omega_B)-X(-\Omega_B)]/2\) for each local observable \(X\). In Fig.~\ref{fig:BTO_profiles}(b), applying this procedure to the local kinetic energy gives \(\Delta_B\mathrm{KE}\). Since local kinetic energy measures the vibrational energy stored in the lattice, this map visualizes the transverse thermal response produced by the gyroscopic term. In a finite sample, the deflection of the phononic heat current appears as a change in the local kinetic-energy profile near the transverse edges that is odd in the magnetic field \cite{Strohm2005,Sheng2006,Kagan2008,Zhang2010PHE,Saito2019PHE}.

The central result of our study is the polarization response shown in Fig.~\ref{fig:BTO_profiles}(c). The gyroscopic coupling deflects the thermally driven lattice modes in the transverse direction, as expected from the phonon Hall effect. In a ferroelectric, the same transverse transport also involves ferrons and therefore produces an accumulation of electric polarization at the transverse edges. The polarization shift \(\Delta_B P\) has opposite signs at the two edges and reverses between the two sides of the hot band, following the reversal of the longitudinal temperature gradient. This odd-in-field and odd-in-gradient edge accumulation of longitudinal ferroelectric polarization is the \textit{ferron Hall effect}. For BaTiO$_3$, the estimated edge accumulation reaches \(16~\mu{\rm C}/{\rm m}^2\), about four orders of magnitude smaller than the equilibrium ferroelectric polarization.

\section{Discussion}

The present work should be viewed as a microscopic demonstration of principle. The magnetic field enters through an effective gyroscopic term that represents the effect of a Zeeman-like coupling between phonon angular momentum and a magnetic field or magnetic order. We treat the coupling strength as a model parameter, but material-specific values must ultimately be obtained from microscopic magnetophononic couplings or be measured experimentally. In real samples, the transverse polarization profile may also be modified by domains and domain walls, surfaces and electrostatic boundary conditions, mobile screening charges, and defects \cite{Catalan2012DomainWallNanoelectronics,Kalinin2018SurfaceScreening,Bulanadi2024DefectsDomainWalls}. Future work should therefore combine material-specific magnetic couplings with realistic ferroelectric domain structures and device geometries.

The ferron Hall effect is fundamentally different from other thermally driven electric responses. In the Seebeck effect, a temperature gradient generates a longitudinal thermoelectric voltage or charge current \cite{HeTritt2017Thermoelectrics}. In pyroelectricity, a change in temperature changes the polarization of a polar material and can generate a current during temperature modulation \cite{Lubomirsky2012PyroelectricGuide}. In thermopolarization, a temperature gradient can induce an electric polarization in an insulator \cite{Onishi2025Thermopolarization}. By contrast, the ferron Hall effect produces opposite polarization shifts at opposite transverse edges and is odd under reversal of both the magnetic field and the longitudinal thermal gradient.

The estimated polarization accumulation in BaTiO$_3$ is approximately \(16~\mu{\rm C}/{\rm m}^2\). This value is much smaller than the spontaneous polarization of conventional ferroelectrics, but comparable to weak polar signals reported in magnetically induced ferroelectrics and related multiferroics, where equilibrium polarizations are on the \(\mu{\rm C}/{\rm m}^2\) scale \cite{Khomskii2009Multiferroics,Lu2019SinglePhaseMultiferroics,Scott2013RoomTemperatureMultiferroics}. This suggests that ferron Hall accumulation is small but well within experimentally measurable magnitudes, especially in materials with strong magnetophononic coupling or optimized edge geometries.

A possible experiment could use a thermal Hall-bar geometry adapted from phonon Hall measurements \cite{Grissonnanche2019,Li2020SrTiO3PHE,Jin2025}: a longitudinal temperature gradient is applied along the sample, an out-of-plane magnetic field is applied perpendicular to it, and the polarization response is measured at the transverse side edges. Combined with recent ferron-transport detection schemes \cite{zhouSurfaceFerronExcitations2023,shenObservationFerronTransport2025,Choe2026Ferrons}, the transverse edge polarization could be probed with piezoresponse force microscopy \cite{Gruverman2019PFM} or with spatially resolved second-harmonic generation \cite{Denev2011SHG}. A complementary electrical measurement could use side electrodes on the transverse edges and modulated heating, with lock-in detection of the component that is odd under both reversal of the applied magnetic field and reversal of the thermal gradient \cite{Bhatia2012PyroelectricCurrent,Yamamoto2021TempModPyroelectric}.

More broadly, our results extend ferronics from longitudinal polarization transport to transverse Hall-type responses. They show that ferrons can play a role analogous to magnons in magnetic insulators, but with electric polarization as the transported quantity. The ferron Hall effect therefore provides a route toward thermal and magnetic control of polarization in ferroelectrics and suggests that excitations in other lattices described by local double-well potentials, such as ferroaxiality and chirality, may support analogous transverse responses.

\section*{Acknowledgments}
We thank Michael Fechner, Mike Pols, Carl Romao, Nicola Spaldin, and Maxim Mostovoy for useful discussions. This research was supported by the ERC Starting Grant CHIRALPHONONICS, no. 101166037.

\bibliography{refs}
\clearrevtexfrontmatternotes

\clearpage
\onecolumngrid
\setcounter{section}{0}
\setcounter{subsection}{0}
\setcounter{subsubsection}{0}
\setcounter{figure}{0}
\setcounter{table}{0}
\setcounter{equation}{0}
\renewcommand{\thesection}{S\arabic{section}}
\renewcommand{\thesubsection}{S\arabic{section}.\arabic{subsection}}
\renewcommand{\thefigure}{S\arabic{figure}}
\renewcommand{\thetable}{S\arabic{table}}
\renewcommand{\theequation}{S\arabic{equation}}
\section*{Supplemental Material for ``Ferron Hall Effect: Transverse Polarization Accumulation Driven by Thermal Gradients in Ferroelectrics''}

\begin{bibunit}[apsrev4-2]
\section{Methodology}

\subsection{Soft-mode description of the ferroelectric lattice}

We consider insulating ferroelectrics whose low-energy lattice dynamics are governed by a polar soft mode of the high-symmetry paraelectric phase. In such materials, the ferroelectric transition is associated with an instability of the centrosymmetric structure toward a lower-symmetry polar distortion. We describe this low-energy sector in terms of a local soft polar mode
\[
\vect{u}_i=(u_{i,x},u_{i,y},u_{i,z}),
\]
defined for each primitive cell $i$. The vector $\vect{u}_i$ is written in a Cartesian representation of the polar basis and specifies the amplitude and orientation of the local distortion relative to the centrosymmetric reference structure. In the effective description used here, $\vect{u}_i$ is the lattice degree of freedom associated with the ferroelectric order parameter.

For a chosen ordered state with polarization direction $\hat{\vect{e}}_{\parallel}$, we decompose the local mode as
\[
\vect{u}_i=u_{i,\parallel}\hat{\vect{e}}_{\parallel}+\vect{u}_{i,\perp},
\qquad
u_{i,\parallel}=\hat{\vect{e}}_{\parallel}\cdot\vect{u}_i.
\]
The local free-energy landscape is described by a double well along the ferroelectric coordinate. The centrosymmetric configuration lies at the unstable saddle, whereas the minima correspond to symmetry-related polar states,
\[
\vect{u}_i=\pm u_0\hat{\vect{e}}_{\parallel},
\]
with $u_0$ the equilibrium amplitude of the local polar distortion. The longitudinal component $u_{i,\parallel}$ changes the amplitude and sign of the distortion along the ferroelectric coordinate and therefore controls the local strength of the order parameter. The transverse component $\vect{u}_{i,\perp}$ describes local reorientation of the polar distortion away from the chosen polar axis. Within a given well, changes in $u_{i,\parallel}$ measure local suppression or enhancement of the ferroelectric distortion, whereas $\vect{u}_{i,\perp}$ describes fluctuations orthogonal to that distortion.

The reduced coordinate $\vect{u}_i$ is constructed from the atomic displacement pattern of the unstable soft mode. Let $\vect{v}_{\alpha,s}$ denote the displacement pattern of atom $s$ associated with the Cartesian soft-mode coordinate $u_{i,\alpha}$, with $\alpha\in\{x,y,z\}$. The displacement of atom $s$ in cell $i$ is then written as
\[
\Delta \vect{r}_{is}=\sum_{\alpha=x,y,z}u_{i,\alpha}\,\vect{v}_{\alpha,s}.
\]
This construction retains the internal displacement pattern of the soft polar instability while reducing the lattice dynamics to one three-component vector degree of freedom per primitive cell. The remaining lattice degrees of freedom enter through the effective interactions acting on $\vect{u}_i$.

Because the soft mode is polar, it carries electric polarization. Because it remains a genuine dynamical lattice coordinate, it also carries kinetic energy and can support internal orbital motion. These observables are obtained from the same projected degree of freedom.

\subsection{Polarization, kinetic energy, and angular momentum}

The local soft mode carries electric polarization through the mode effective charge tensor $\matr{Z}$, obtained by projecting the Born effective charges onto the soft-mode basis,
\[
(\matr{Z})_{\mu\alpha}
=
\sum_s\sum_{\nu}(\matr{Z}_s^*)_{\mu\nu}(\vect{v}_{\alpha,s})_{\nu},
\]
where $\mu,\nu\in\{x,y,z\}$ are Cartesian indices and $\matr{Z}_s^*$ is the Born effective charge tensor of atom $s$. The dipole moment of unit cell $i$ is
\[
\vect{p}_i=\matr{Z}\vect{u}_i,
\]
so that the local polarization is
\[
\vect{P}_i=\frac{1}{\Omega}\matr{Z}\vect{u}_i,
\]
with $\Omega$ the primitive-cell volume. The local soft mode is thus the polarization-carrying degree of freedom of the effective theory.

The same coordinate also carries kinetic energy. The projected mode-mass matrix is
\[
\matr{M}_{\alpha\beta}=\sum_s m_s\,\vect{v}_{\alpha,s}\cdot\vect{v}_{\beta,s},
\]
and the local kinetic energy in unit cell $i$ is
\[
E_{\mathrm{kin},i}=\frac{1}{2}\dot{\vect{u}}_i^{\,T}\matr{M}\dot{\vect{u}}_i.
\]
The matrix $\matr{M}$ contains the inertia associated with the soft-mode coordinates after projection from the full atomic degrees of freedom.

The projected dynamics also supports local phonon angular momentum. To remove rigid translation of the primitive cell, the mode basis is measured relative to the cell center of mass,
\[
\tilde{\vect{v}}_{\alpha,s}=\vect{v}_{\alpha,s}-\vect{v}_{\alpha}^{\mathrm{cm}},
\qquad
\vect{v}_{\alpha}^{\mathrm{cm}}=\frac{1}{M_{\mathrm{cell}}}\sum_s m_s\vect{v}_{\alpha,s},
\]
where $M_{\mathrm{cell}}=\sum_s m_s$. The local angular momentum of unit cell $i$ for Cartesian component $k\in\{x,y,z\}$ is then written as
\[
L_{i,k}=\vect{u}_i^T\matr{\Lambda}^{(k)}\dot{\vect{u}}_i,
\]
with
\[
(\matr{\Lambda}^{(k)})_{\alpha\beta}
=
\sum_s m_s\,[\tilde{\vect{v}}_{\alpha,s}\times\tilde{\vect{v}}_{\beta,s}]_k.
\]
The matrices $\matr{\Lambda}^{(k)}$ encode the internal orbital motion of the projected soft-mode dynamics. Within this description, the same local field $\vect{u}_i$ controls the polarization channel through $\matr{Z}$, the energy channel through $\matr{M}$, and the angular-momentum channel through $\matr{\Lambda}^{(k)}$.

\subsection{Effective Hamiltonian}

The energy-conserving part of the projected local-mode dynamics is described by
\begin{equation}
\mathcal{H}=E_{\rm kin}+V_{\rm on}+V_{\rm sr}+V_{\rm dip},
\end{equation}
where \(E_{\rm kin}=\sum_i E_{{\rm kin},i}\). The potential energy \(V=V_{\rm on}+V_{\rm sr}+V_{\rm dip}\) contains the local anharmonic ferroelectric instability, short-range intercell couplings, and the long-range dipole-dipole interaction. The Hamiltonian forces are obtained from
\begin{equation}
\vect F_i=-\frac{\partial V}{\partial \vect u_i}.
\end{equation}

\subsection{Equation of motion}

The dynamics of the projected soft mode is governed by
\begin{equation}
\matr M\ddot{\vect u}_i
=
\vect F_i^{\rm on}
+
\vect F_i^{\rm sr}
+
\vect F_i^{\rm dip}
+
\matr G_{\Omega_B}\dot{\vect u}_i
-
\gamma_i\matr M\dot{\vect u}_i
+
\vect\eta_i(t).
\end{equation}
Here \(\matr M\) is the projected mode-mass matrix, \(\vect F_i^{\rm on}\) is the on-site force, \(\vect F_i^{\rm sr}\) is the short-range intercell force, \(\vect F_i^{\rm dip}\) is the long-range dipole-dipole force, \(\matr G_{\Omega_B}\dot{\vect u}_i\) is the gyroscopic force, \(\gamma_i\) is the local damping coefficient, and \(\vect\eta_i(t)\) is the stochastic force associated with the local Langevin bath. This equation describes the nonequilibrium dynamics of the polarization-carrying soft mode in the presence of thermal driving and broken time-reversal symmetry.

\subsection{On-site potential and local restoring force}

The on-site potential is written in terms of the longitudinal coordinate and the transverse distortion as
\begin{equation}
V_{\rm on}
=
\sum_i
\left[
\frac{A_2}{2}u_{i,\parallel}^2
+
\frac{B_4}{4}u_{i,\parallel}^4
+
\frac{C_6}{6}u_{i,\parallel}^6
+
\frac{\kappa_\perp}{2}|\vect u_{i,\perp}|^2
\right].
\end{equation}
The longitudinal part describes the double-well energy landscape along the ferroelectric coordinate, while the transverse term provides harmonic confinement away from the chosen polar axis. The corresponding force is
\begin{equation}
\vect F_i^{\rm on}
=
-\left(A_2+B_4u_{i,\parallel}^2+C_6u_{i,\parallel}^4\right)u_{i,\parallel}\hat{\vect e}_{\parallel}
-\kappa_\perp\vect u_{i,\perp}.
\end{equation}
The longitudinal coefficients \(A_2\), \(B_4\), and \(C_6\) are obtained from the projected double-well energy landscape, while \(\kappa_\perp\) is chosen as a weak positive transverse confinement parameter.

\subsection{Short-range intercell interactions}

Short-range intercell couplings are obtained by projecting the first-principles interatomic force constants onto the soft-mode basis. If \(\mgreek{\Phi}_{st}(\vect R)\) denotes the interatomic force-constant block coupling atom \(t\) in one primitive cell to atom \(s\) in a cell displaced by lattice vector \(\vect R\), then the projected short-range kernel is
\begin{equation}
(\matr K^{\rm sr})_{\alpha\beta}(\vect R)
=
\sum_{s,t}
\vect v_{\alpha,s}^{\,T}
\mgreek{\Phi}_{st}(\vect R)
\vect v_{\beta,t}.
\end{equation}
With the zero-sum short-range kernel used in the simulations, the short-range intercell energy is written as
\begin{equation}
V_{\rm sr}
=
\frac{1}{2}
\sum_i
\sum_{\vect R}
\vect u_i^{\,T}
\matr K^{\rm sr}(\vect R)
\vect u_{i+\vect R}.
\end{equation}
Here \(\sum_{\vect R}\matr K^{\rm sr}(\vect R)\simeq 0\). The corresponding force is equivalently written in pair-difference form,
\begin{equation}
\vect F_i^{\rm sr}
=
-\sum_{\vect R}
\matr K^{\rm sr}(\vect R)
\left(\vect u_{i+\vect R}-\vect u_i\right).
\end{equation}
This term represents the short-range energetic cost of relative polar distortions between unit cells. For open boundaries, bonds that would connect outside the physical simulation region are omitted; bonds crossing periodic boundaries are wrapped periodically.

\subsection{Long-range dipolar interaction}

The long-range electrostatic interaction is generated by the dipole moment carried by the local soft mode. In the simulations, the dipolar force is evaluated in reciprocal space using fast Fourier transforms. For periodic directions, the Fourier transform is taken over the physical simulation length. For open directions, the physical sample is placed inside a larger real-space array with \(\vect u=\vect 0\) on the added sites outside the sample. After transforming back to real space, only the force on the physical sample is retained.

Using the Fourier convention on this real-space array,
\begin{equation}
\vect u(\vect q)
=
\sum_i
\vect u_i
e^{-i\vect q\cdot\vect R_i},
\qquad
\vect u_i
=
\frac{1}{N_{\rm FT}}
\sum_{\vect q}
\vect u(\vect q)
e^{i\vect q\cdot\vect R_i},
\end{equation}
where the sum over \(i\) includes the physical sites and the added zero-\(\vect u\) sites. Here \(N_{\rm FT}\) is the total number of sites used in the Fourier transform. The dipolar energy is
\begin{equation}
V_{\rm dip}
=
\frac{1}{2N_{\rm FT}}
\sum_{\vect q\neq\vect 0}
\vect u(-\vect q)^{T}
\matr K^{\rm dip}(\vect q)
\vect u(\vect q),
\end{equation}
with
\begin{equation}
\matr K^{\rm dip}(\vect q)
=
\frac{4\pi k_e}{\Omega}
\frac{
\left(\matr Z^{T}\vect q\right)
\left(\matr Z^{T}\vect q\right)^{T}
}{
\vect q^{\,T}\mgreek{\epsilon}_{\infty}\vect q
}.
\end{equation}
Here \(k_e\) is the Coulomb constant and \(\mgreek{\epsilon}_{\infty}\) is the electronic dielectric tensor. The kernel describes the screened Coulomb energy associated with the longitudinal component of the mode-induced dipole density. The corresponding reciprocal-space force is
\begin{equation}
\vect F^{\rm dip}(\vect q)
=
-
\matr K^{\rm dip}(\vect q)\vect u(\vect q),
\qquad
\vect F^{\rm dip}(\vect q=\vect 0)=\vect 0.
\end{equation}
The \(\vect q=\vect 0\) component corresponds to a spatially uniform macroscopic polarization, whose electrostatic contribution depends on electrical boundary conditions. We set this component to zero, corresponding to no imposed macroscopic depolarizing field, and retain the finite-\(\vect q\) depolarizing fields associated with polarization inhomogeneities.

For the slab geometry used here, the simulations are periodic along \(x\) and \(z\), while the transverse \(y\) direction is finite. The real-space array used for the Fourier transform is therefore enlarged only along \(y\). The added zero-\(\vect u\) region separates the physical slab from its artificial transverse periodic images and reduces image interactions while retaining the reciprocal-space evaluation of the long-range dipolar force.

\subsection{Magnetic-field coupling}

The coupling to the external magnetic field is introduced through an effective gyroscopic term acting on the soft-mode velocity,
\[
\matr{G}_{\Omega_B}
=\Omega_B
\matr{M}^{1/2}\matr{C}(\hat{\vect{B}})\matr{M}^{1/2},
\]
where $\Omega_B$ is an effective coupling rate and $\hat{\vect{B}}$ is the field direction. The antisymmetric generator $\matr{C}(\hat{\vect{B}})$ is
\[
\matr{C}(\hat{\vect{B}})
=
\begin{pmatrix}
0 & -\hat{B}_z & \hat{B}_y\\
\hat{B}_z & 0 & -\hat{B}_x\\
-\hat{B}_y & \hat{B}_x & 0
\end{pmatrix}.
\]
The term $\matr{G}_{\Omega_B}\dot{\vect{u}}_i$ mixes the velocity components through an antisymmetric rotation in the projected mode space and does not modify the potential energy landscape. In the driven nonequilibrium state, this term produces the field-odd transverse redistribution of the soft-mode dynamics.

\subsection{Langevin driving and thermal profile}

Nonequilibrium driving is introduced through a spatially varying Langevin bath. The damping coefficient $\gamma_i$ and the local temperature $T_i$ are allowed to vary from cell to cell, so that the system can be driven by a prescribed thermal profile while stronger damping is maintained in selected boundary regions. The stochastic force satisfies
\[
\langle\vect{\eta}_i(t)\rangle=\vect{0},
\]
and
\[
\langle\eta_{i,\alpha}(t)\eta_{j,\beta}(t')\rangle
=
2\gamma_i k_B T_i\,\matr{M}_{\alpha\beta}\,\delta_{ij}\delta(t-t').
\]
The noise correlator is therefore consistent with the projected mass matrix. The temperature profile used in the simulations consists of a hot central region and colder outer regions, producing oppositely directed longitudinal temperature gradients on the two sides of the sample. This geometry makes it possible to resolve the transverse response directly in real space.

\subsection{Time integration}

The equations of motion are integrated by splitting each time step into conservative, gyroscopic, and Langevin parts. The conservative forces are applied through half-step velocity updates. The gyroscopic part is solved exactly over each substep as a rotation in velocity space generated by the antisymmetric magnetic coupling. The damping and stochastic terms are propagated through the exact Ornstein--Uhlenbeck update associated with the Langevin bath. The full integrator is assembled in symmetric BAOAB form.

\subsection{First-principles parametrization and validation of BaTiO$_3$}
\label{sec:bto_first_principles_validation}

We construct the BaTiO$_3$ local-mode model from first-principles structural, dielectric, and lattice-dynamical calculations. The model used in the nonequilibrium simulations is a cubic-reference local-mode Hamiltonian: the three Cartesian local-mode coordinates are defined from the triply degenerate unstable $\Gamma$-point polar subspace of cubic paraelectric BaTiO$_3$, while the onsite anharmonic potential is calibrated from a fixed-cell DFT switching path connecting the two symmetry-related tetragonal ferroelectric variants. This construction follows the first-principles effective-Hamiltonian description of perovskite ferroelectrics, in which the low-energy polar distortions are expanded around the high-symmetry cubic reference and the anharmonic energy surface stabilizes the polar wells~\cite{Zhong1994PRL_BaTiO3,Zhong1995PRB_BaTiO3}. The use of a cubic soft-mode basis is natural for BaTiO$_3$ because the cubic phase has a threefold-degenerate polar instability whose condensation generates the low-symmetry ferroelectric phases.

All density-functional-theory calculations were performed using the Vienna \textit{ab initio} simulation package (\textsc{VASP})~\cite{Kresse1996PRB_VASP,Kresse1999PRB_PAW} with projector-augmented-wave (PAW) potentials. We used the PAW datasets Ba\_sv, Ti\_pv, and O, corresponding to valence configurations Ba $(5s^2 5p^6 6s^2)$, Ti $(3p^6 3d^3 4s^1)$, and O $(2s^2 2p^4)$. Exchange and correlation were treated with the PBEsol generalized-gradient approximation~\cite{Perdew2008PRL_PBEsol}. A plane-wave energy cutoff of 600~eV was used throughout, together with \texttt{PREC = Accurate}, \texttt{LREAL = .FALSE.}, \texttt{LASPH = .TRUE.}, and \texttt{ADDGRID = .TRUE.}. Primitive-cell structural relaxations used a $\Gamma$-centered $12\times12\times12$ $k$-point mesh, Gaussian smearing with \texttt{ISMEAR = 0} and $\sigma=0.01$~eV, and electronic convergence thresholds of $10^{-9}$--$10^{-10}$~eV. Ionic relaxations were continued until the Hellmann--Feynman forces were below $10^{-4}$~eV/\AA\ and the residual stresses were below $10^{-2}$~kbar.

Harmonic interatomic force constants were computed using the frozen-phonon method as implemented in \textsc{Phonopy}~\cite{Togo2015ScrMater_Phonopy,Togo2023JPSJ_Phonopy}. We used $3\times3\times3$ supercells containing 135 atoms and finite displacements of 0.01~\AA. The supercell force calculations used the same PAW datasets, exchange-correlation functional, cutoff, and electronic convergence settings as the primitive-cell calculations, with a $\Gamma$-centered $4\times4\times4$ $k$-point mesh. Born effective charges and electronic dielectric tensors were obtained from linear-response calculations in \textsc{VASP}. Non-analytical corrections (NACs) for LO--TO splitting were evaluated in \textsc{Phonopy} using the calculated Born effective charges and electronic dielectric tensors. The force-constant and response calculations were carried out for both cubic and tetragonal structures for validation, although the simulation model uses the cubic force constants, cubic Born charges, and cubic electronic dielectric tensor as its reference parameters.

\begin{figure*}[tbh!]
    \centering
    \includegraphics[width=0.99\textwidth]{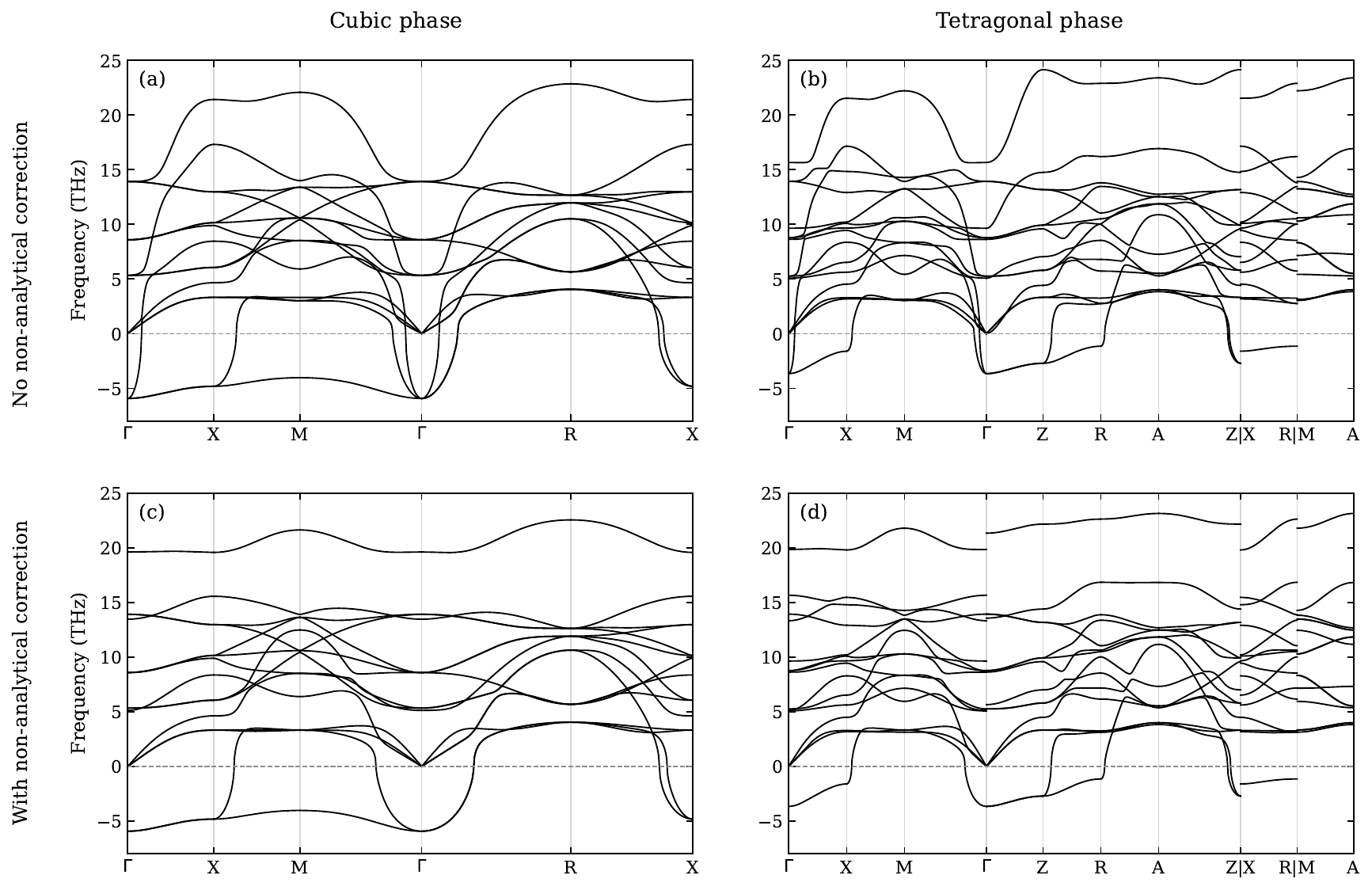}
    \caption{
    Phonon dispersions of cubic and tetragonal BaTiO$_3$ computed from first-principles finite-displacement interatomic force constants using $3\times3\times3$ supercells. The left column shows the cubic phase and the right column shows the tetragonal phase. The upper row shows the dispersions obtained without the non-analytical correction (NAC), while the lower row includes the NAC/LO--TO correction using the calculated Born effective charges and electronic dielectric tensor. The comparison between the uncorrected and corrected spectra highlights the large polar LO--TO splitting characteristic of BaTiO$_3$. The overall frequency range and unstable low-frequency branches are consistent with recent first-principles phonon calculations for cubic BaTiO$_3$~\cite{Cazorla2025OpticalControl}; computed phonon data for tetragonal BaTiO$_3$ can also be inspected in the Materials Project entry mp-5986~\cite{MaterialsProject2013,MaterialsProjectBaTiO3mp5986}.}
    \label{fig:bto_phonon_dispersion_3x3}
\end{figure*}

\begin{table*}[t]
\centering
\caption{Summary of first-principles settings used to parametrize the BaTiO$_3$ local-mode model.}
\label{tab:bto_dft_setup}
\renewcommand{\arraystretch}{1.18}
\begin{tabular}{p{0.27\textwidth} p{0.24\textwidth} p{0.40\textwidth}}
\hline\hline
Quantity & Value & Purpose \\
\hline
Exchange-correlation functional
& PBEsol~\cite{Perdew2008PRL_PBEsol}
& Structural, response, and phonon calculations \\

PAW datasets
& Ba\_sv, Ti\_pv, O
& Ba and Ti semicore states included \\

Plane-wave cutoff
& 600~eV
& Used for all VASP calculations \\

Primitive-cell $k$ mesh
& $12\times12\times12$ $\Gamma$ centered
& Relaxations, static energies, response, and Berry-phase path \\

Phonon supercell
& $3\times3\times3$, 135 atoms
& Finite-displacement force constants \\

Supercell $k$ mesh
& $4\times4\times4$ $\Gamma$ centered
& Frozen-phonon force calculations \\

Finite displacement
& 0.01~\AA
& Phonopy force-constant extraction \\

Electronic convergence
& $10^{-9}$--$10^{-10}$~eV
& Relaxation, static, response, and path calculations \\

Force threshold
& $10^{-4}$~eV/\AA
& Primitive-cell relaxations \\

Residual stress threshold
& $10^{-2}$~kbar
& Primitive-cell relaxations \\

NAC input
& $Z^*$ and $\epsilon_\infty$ from linear response
& LO--TO splitting and dipolar model \\
\hline\hline
\end{tabular}
\end{table*}

The relaxed structural parameters are summarized in Table~\ref{tab:bto_structure_validation}. The cubic lattice constant agrees with PBEsol benchmarks for stress-free BaTiO$_3$, and the tetragonal structure lies in the expected zero-temperature first-principles range. The tetragonality and polarization are larger than room-temperature experimental values, but this is expected for a zero-temperature DFT calculation and is consistent with previous PBEsol assessments, where comparison to low-temperature or Ginzburg--Landau-extrapolated criteria gives a more appropriate validation than direct comparison to room-temperature data~\cite{Kwei1993JPC_BTO,Watanabe2018JCP_BTO}. The relaxed tetragonal phase is lower than the relaxed cubic reference by 20.056~meV/f.u. This energy difference should not be confused with the fixed-cell coherent switching barrier used below to calibrate the onsite potential.

\begin{table*}[t]
\centering
\caption{Structural and polarization validation for BaTiO$_3$. Energies are per five-atom formula unit. The model polarization is obtained from the cubic local-mode charge and projected well amplitude.}
\label{tab:bto_structure_validation}
\resizebox{\textwidth}{!}{
\begin{tabular}{llcccccl}
\hline\hline
Source & Phase/model & $a$ (\AA) & $c$ (\AA) & $c/a$ & $\Omega$ (\AA$^3$) & $P_s$ (C/m$^2$) & Comment \\
\hline
This work & Cubic & 3.985236 & 3.985236 & 1.000000 & 63.293954 & -- & PBEsol relaxed reference \\
This work & Tetragonal & 3.969504 & 4.066228 & 1.024367 & 64.071393 & 0.346588 & PBEsol relaxed endpoint used in fixed-cell Berry path \\
This work & Tetragonal model & -- & -- & -- & 63.293954 & 0.3582 & $(\partial P/\partial u)u_0$ from cubic local mode \\
Ref.~\cite{Watanabe2018JCP_BTO} & Cubic & 3.986 & 3.986 & 1.000 & 63.31 & 0 & PBEsol stress-free \\
Ref.~\cite{Watanabe2018JCP_BTO} & Tetragonal & 3.971 & 4.059 & 1.022 & 64.01 & 0.320 & PBEsol stress-free \\
Refs.~\cite{Kwei1993JPC_BTO,Watanabe2018JCP_BTO} & Tetragonal & 3.993 & 4.037 & 1.011 & 64.35 & 0.254 & Room-temperature experiment \\
Ref.~\cite{Watanabe2018JCP_BTO} & Tetragonal & 3.992 & 4.051 & 1.015 & 64.56 & 0.318 & Approximate 0~K GL/extrapolated criterion \\
\hline\hline
\end{tabular}}
\end{table*}

The dielectric tensors and Born effective charges are given in Tables~\ref{tab:bto_dielectric_validation} and~\ref{tab:bto_born_validation}. The anomalously large cubic Ti and longitudinal oxygen Born charges reproduce the established dynamical-charge pattern of BaTiO$_3$~\cite{Ghosez1995PRB_BTOBorn,Ghosez1998DynamicalCharges,Zhong1994Zstar}. In the tetragonal phase, the reduction of $Z^*_{\mathrm{Ti},zz}$ and of the polar-chain oxygen charge is accompanied by a smaller $\epsilon_{\infty,zz}$, consistent with the reduced dynamical charge transfer along the ferroelectric Ti--O chain~\cite{Ghosez1998DynamicalCharges}. The ionic and total dielectric entries are reported as response-calculation diagnostics; the simulations use only $\epsilon_\infty$ for electronic screening of the explicit dipole-dipole interaction. In particular, the static lattice dielectric response of the unstable cubic reference is not used as a physical screening parameter.

\begin{table*}[t]
\centering
\caption{Electronic, ionic, and total dielectric tensors from linear response. Only diagonal components are shown; off-diagonal components are zero within numerical precision. The cubic $\epsilon_\infty$ tensor is used in the screened dipole-dipole kernel.}
\label{tab:bto_dielectric_validation}
\begin{tabular}{llccc}
\hline\hline
Phase & Tensor & $xx$ & $yy$ & $zz$ \\
\hline
Cubic & $\epsilon_\infty$ & 6.869406 & 6.869406 & 6.869406 \\
Cubic & $\epsilon_\mathrm{ion}$ & 0.564635 & 0.564635 & 0.564635 \\
Cubic & $\epsilon_\infty+\epsilon_\mathrm{ion}$ & 7.434041 & 7.434041 & 7.434041 \\
\hline
Tetragonal & $\epsilon_\infty$ & 6.544050 & 6.544050 & 5.743872 \\
Tetragonal & $\epsilon_\mathrm{ion}$ & 0.666429 & 0.666429 & 20.448042 \\
Tetragonal & $\epsilon_\infty+\epsilon_\mathrm{ion}$ & 7.210479 & 7.210479 & 26.191914 \\
\hline\hline
\end{tabular}
\end{table*}

\begin{table*}[t]
\centering
\caption{Born effective charge validation. For cubic BaTiO$_3$, $O_\parallel$ denotes the component along the Ti--O bond and $O_\perp$ the two transverse components. For tetragonal BaTiO$_3$, the polar axis is $z$ and O$_2$ is the oxygen on the polar Ti--O chain.}
\label{tab:bto_born_validation}
\resizebox{\textwidth}{!}{
\begin{tabular}{llcccl}
\hline\hline
Source & Phase/atom & $Z^*_{xx}$ & $Z^*_{yy}$ & $Z^*_{zz}$ & Comment \\
\hline
This work & Cubic Ba & 2.74964 & 2.74964 & 2.74964 & Isotropic \\
This work & Cubic Ti & 7.46679 & 7.46679 & 7.46679 & Isotropic \\
This work & Cubic O$_1$ & $-2.14870$ & $-5.91902$ & $-2.14870$ & $O_\parallel$ along $y$ \\
This work & Cubic O$_2$ & $-5.91902$ & $-2.14870$ & $-2.14870$ & $O_\parallel$ along $x$ \\
This work & Cubic O$_3$ & $-2.14870$ & $-2.14870$ & $-5.91902$ & $O_\parallel$ along $z$ \\
Ref.~\cite{Ghosez1998DynamicalCharges} & Cubic Ba & 2.77 & 2.77 & 2.77 & Linear response \\
Ref.~\cite{Ghosez1998DynamicalCharges} & Cubic Ti & 7.25 & 7.25 & 7.25 & Linear response \\
Ref.~\cite{Ghosez1998DynamicalCharges} & Cubic $O_\perp$ & $-2.15$ & $-2.15$ & $-2.15$ & Transverse oxygen charge \\
Ref.~\cite{Ghosez1998DynamicalCharges} & Cubic $O_\parallel$ & $-5.71$ & $-5.71$ & $-5.71$ & Longitudinal oxygen charge \\
\hline
This work & Tetragonal Ba & 2.72849 & 2.72849 & 2.83457 & Polar axis $z$ \\
This work & Tetragonal Ti & 7.12307 & 7.12307 & 5.69244 & Reduced along $z$ \\
This work & Tetragonal O$_1$ & $-5.70537$ & $-2.13596$ & $-1.93832$ & Transverse oxygen \\
This work & Tetragonal O$_2$ & $-2.01023$ & $-2.01023$ & $-4.65037$ & Oxygen on polar chain \\
This work & Tetragonal O$_3$ & $-2.13596$ & $-5.70537$ & $-1.93832$ & Transverse oxygen \\
Ref.~\cite{Ghosez1998DynamicalCharges} & Tetragonal Ba & 2.72 & 2.72 & 2.83 & Experimental tetragonal structure \\
Ref.~\cite{Ghosez1998DynamicalCharges} & Tetragonal Ti & 6.94 & 6.94 & 5.81 & Experimental tetragonal structure \\
Ref.~\cite{Ghosez1998DynamicalCharges} & Tetragonal $O_\parallel$ & $-1.99$ & $-1.99$ & $-4.73$ & Oxygen on polar chain \\
Ref.~\cite{Ghosez1998DynamicalCharges} & Tetragonal $O_\perp$ & $-2.14$/$-5.53$ & $-5.53$/$-2.14$ & $-1.95$ & Transverse oxygens \\
\hline\hline
\end{tabular}}
\end{table*}

The cubic $\Gamma$-point phonon spectrum is summarized in Table~\ref{tab:bto_cubic_gamma_validation}. Without the non-analytical correction, the cubic phase has a triply degenerate unstable polar soft mode at $-5.933489$~THz, corresponding to $197.920i$~cm$^{-1}$. For $\mathbf q\parallel z$, the non-analytical Coulomb term leaves the two transverse soft branches unstable and shifts the longitudinal polar branches upward, producing the characteristic LO--TO splitting of perovskite ferroelectrics~\cite{Zhong1994Zstar}.

\begin{table*}[t]
\centering
\caption{Cubic BaTiO$_3$ zone-center phonon frequencies from the present $3\times3\times3$ finite-displacement force constants. Frequencies are given in THz, with imaginary modes shown as negative values. NAC denotes the non-analytical LO--TO correction for $\mathbf q\parallel z$.}
\label{tab:bto_cubic_gamma_validation}
\begin{tabular}{rcc}
\hline\hline
Mode & No NAC & NAC $\mathbf q\parallel z$ \\
\hline
1  & $-5.933489$ & $-5.933489$ \\
2  & $-5.933489$ & $-5.933489$ \\
3  & $-5.933489$ & $0.000000$ \\
4  & $0.000000$ & $0.000000$ \\
5  & $0.000000$ &  $0.000000$ \\
6  & $0.000000$ &  $5.119519$ \\
7  &  $5.330637$ &  $5.330637$ \\
8  &  $5.330637$ &  $5.330637$ \\
9  &  $5.330637$ &  $8.581325$ \\
10 &  $8.581325$ &  $8.581325$ \\
11 &  $8.581325$ &  $8.581325$ \\
12 &  $8.581325$ & $13.465355$ \\
13 & $13.916358$ & $13.916358$ \\
14 & $13.916358$ & $13.916358$ \\
15 & $13.916358$ & $19.617673$ \\
\hline\hline
\end{tabular}
\end{table*}

The cubic IR-active frequencies are compared with first-principles and experimental LO--TO benchmarks in Table~\ref{tab:bto_cubic_loto_comparison}. The calculated TO$_2$, TO$_3$, LO$_1$, and LO$_2$ frequencies agree closely with Ref.~\cite{Zhong1994Zstar}. The soft TO$_1$ branch is slightly more unstable, while LO$_3$ is lower than the benchmark value. This pattern is consistent with the sensitivity of the highest longitudinal branch to equilibrium volume, exchange-correlation functional, electronic dielectric screening, and the non-analytic Coulomb term. The calculated spectrum nevertheless reproduces the central physical feature needed for the local-mode model: anomalous Born charges generate a large polar LO--TO splitting with the correct mode ordering.

\begin{table*}[t]
\centering
\caption{Cubic BaTiO$_3$ IR-active zone-center phonons compared with first-principles and experimental LO--TO benchmarks. Frequencies are in cm$^{-1}$; imaginary frequencies are denoted by $i$. The present values are obtained from Table~\ref{tab:bto_cubic_gamma_validation} using $1~\mathrm{THz}=33.35641~\mathrm{cm}^{-1}$.}
\label{tab:bto_cubic_loto_comparison}
\begin{tabular}{lccc}
\hline\hline
Mode & This work & Ref.~\cite{Zhong1994Zstar}, theory & Ref.~\cite{Zhong1994Zstar}, experiment \\
\hline
TO$_1$ & $197.9i$ & $178i$ & -- \\
TO$_2$ & 177.8 & 177 & 181 \\
TO$_3$ & 464.2 & 468 & 487 \\
LO$_1$ & 170.8 & 173 & 180 \\
LO$_2$ & 449.2 & 453 & 468 \\
LO$_3$ & 654.4 & 738 & 717 \\
\hline\hline
\end{tabular}
\end{table*}

The soft-mode eigenvector used in the local-mode model is the rotated $z$-polar member of the degenerate cubic TO$_1$ subspace, normalized by $v_z(\mathrm{Ti})=1$. This Slater-like pattern is dominated by Ti motion against oxygen motion, with the oxygen on the polar Ti--O chain having the largest opposite displacement. The two other basis vectors $v_x$ and $v_y$ are obtained by cubic symmetry. Table~\ref{tab:bto_softmode_eigenvector} lists the local-mode eigenvector and its contribution to the $z$ component of the mode effective charge. The same anomalous Ti and longitudinal-oxygen components that dominate the eigenvector also dominate the mode charge.

\begin{table*}[t]
\centering
\caption{Cubic $z$-polar soft-mode eigenvector used as the local-mode basis and decomposition of the mode effective charge. The eigenvector is normalized by $v_z(\mathrm{Ti})=1$. For the $z$-polar member, O$_1$ and O$_2$ are transverse oxygens and O$_3$ is the oxygen on the polar Ti--O chain.}
\label{tab:bto_softmode_eigenvector}
\begin{tabular}{lccc}
\hline\hline
Atom & Role in $z$-polar mode & $v_z$ & Contribution to $Z_{zz}$ ($e$) \\
\hline
Ba & A-site ion & 0.007454 & 0.0205 \\
Ti & B-site ion & 1.000000 & 7.4668 \\
O$_1$ & $O_\perp$ & $-0.770274$ & 1.6551 \\
O$_2$ & $O_\perp$ & $-0.770274$ & 1.6551 \\
O$_3$ & $O_\parallel$ & $-1.515269$ & 8.9689 \\
\hline
Total & -- & -- & 19.7664 \\
\hline\hline
\end{tabular}
\end{table*}

For the tetragonal structure, the $\Gamma$ spectrum is shown in Table~\ref{tab:bto_tetragonal_gamma_validation}. The twofold transverse instability at $-3.660342$~THz, or $122.096i$~cm$^{-1}$, reflects the zero-temperature tendency of tetragonal BaTiO$_3$ to distort toward lower-symmetry ferroelectric phases~\cite{Zhong1994PRL_BaTiO3,Zhong1995PRB_BaTiO3,Gigli2022npj_BTO}. The non-analytical correction produces strong directional LO--TO splittings, with the highest longitudinal mode reaching 21.337278~THz for $\mathbf q\parallel z$ and 19.845816~THz for $\mathbf q\parallel x$.

\begin{table*}[t]
\centering
\caption{Tetragonal BaTiO$_3$ zone-center phonon frequencies from the present $3\times3\times3$ finite-displacement force constants. Frequencies are given in THz, with imaginary modes shown as negative values. NAC spectra are shown for $\mathbf q\parallel z$ and $\mathbf q\parallel x$ to display the anisotropic LO--TO splitting of the tetragonal phase.}
\label{tab:bto_tetragonal_gamma_validation}
\begin{tabular}{rccc}
\hline\hline
Mode & No NAC & NAC $\mathbf q\parallel z$ & NAC $\mathbf q\parallel x$ \\
\hline
1  & $-3.660342$ & $-3.660342$ & $-3.660342$ \\
2  & $-3.660342$ & $-3.660342$ &  $0.000000$ \\
3  & $0.000000$ & $0.000000$ &  $0.000000$ \\
4  &  $0.000000$ &  $0.000000$ &  $0.000000$ \\
5  &  $0.000000$ &  $0.000000$ &  $5.071924$ \\
6  &  $5.071924$ &  $5.258515$ &  $5.120099$ \\
7  &  $5.258515$ &  $5.258515$ &  $5.258515$ \\
8  &  $5.258515$ &  $5.649918$ &  $8.623348$ \\
9  &  $8.623348$ &  $8.623348$ &  $8.739230$ \\
10 &  $8.764797$ &  $8.764797$ &  $8.764797$ \\
11 &  $8.764797$ &  $8.764797$ &  $9.638163$ \\
12 &  $9.638163$ & $13.562975$ & $13.327337$ \\
13 & $13.934029$ & $13.934029$ & $13.934029$ \\
14 & $13.934029$ & $13.934029$ & $15.661683$ \\
15 & $15.661683$ & $21.337278$ & $19.845816$ \\
\hline\hline
\end{tabular}
\end{table*}

Projecting the cubic Born charges onto the rotated cubic soft-mode basis gives the nearly isotropic mode effective charge tensor
\begin{equation}
Z =
\begin{pmatrix}
19.76637 & 0 & 0 \\
0 & 19.76637 & 0 \\
0 & 0 & 19.76637
\end{pmatrix} e ,
\end{equation}
and hence
\begin{equation}
\frac{\partial P}{\partial u}=\frac{1}{\Omega}Z=5.00351\,I~\mathrm{C\,m^{-2}\,\AA^{-1}},
\end{equation}
where $\Omega=63.293954$~\AA$^3$ is the cubic primitive-cell volume. The large local-mode charge is dominated by the anomalous Ti and longitudinal oxygen Born charges and is the microscopic origin of the strong dipole-dipole interaction in the simulation model.

The onsite double-well potential is calibrated from a fixed-cell DFT switching path between the two tetragonal ferroelectric variants. The path is parameterized by a dimensionless coordinate $\lambda$, with $\lambda=\pm1$ corresponding to the two opposite tetragonal polar variants and $\lambda=0$ to the centrosymmetric midpoint. The DFT energy profile is symmetric to numerical precision and gives
\begin{equation}
\Delta E_b \equiv E(\lambda=0)-E(\lambda=\pm1)=38.43446~\mathrm{meV/f.u.}.
\end{equation}
The Berry-phase polarization of the positive endpoint is
\begin{equation}
P_z(\lambda=+1)=0.346588~\mathrm{C\,m^{-2}},
\end{equation}
with a polarization quantum of 1.016806~C~m$^{-2}$ along $z$. The corresponding cubic local-mode estimate is $(\partial P/\partial u)u_0=0.3582$~C~m$^{-2}$, only 3.35\% larger than the Berry-phase value. The switching path and Berry-phase polarization are written for a tetragonal variant polarized along $z$. In the nonequilibrium simulations, the cubic local-mode parameters can be rotated so that the ordered ferroelectric axis is the simulation axis.

The fixed-cell barrier is close to the PBEsol one-dimensional DFT double-well value of 34.63~meV reported in Ref.~\cite{Esswein2022PRResearch_BTO}. The same reference gives 104.8~meV for PBE, showing the strong dependence of the double-well depth on functional, reference structure, and collective-coordinate definition. The present value is therefore used as the homogeneous fixed-cell tetragonal switching barrier for the onsite potential, not as a macroscopic switching barrier controlled by domain nucleation and domain-wall motion.

For the molecular-dynamics simulations we use a stable quartic onsite potential,
\begin{equation}
V_\mathrm{on}(u_\parallel)=\frac{A_2}{2}u_\parallel^2+\frac{B_4}{4}u_\parallel^4+\frac{C_6}{6}u_\parallel^6,
\end{equation}
with $C_6=0$. The quartic coefficients are chosen to reproduce exactly the projected equilibrium well amplitude and the DFT barrier. A sixth-order least-squares fit was also tested for interpolation of the DFT path, but the unconstrained fit gives a negative sextic coefficient and is unbounded at large amplitude. We therefore use the stable barrier-matched quartic form in the dynamics. The transverse confinement $\kappa_\perp$ is not fitted to a separate first-principles transverse path; it is a weak positive simulation parameter used to confine fluctuations away from the chosen ferroelectric axis.

\begin{table*}[t]
\centering
\caption{Local-mode and onsite-potential parameters used in the nonequilibrium simulations. The local-mode amplitude $u_0$ is obtained by projecting the fixed-cell tetragonal switching displacement onto the cubic soft-mode basis.}
\label{tab:bto_local_model_validation}
\resizebox{\textwidth}{!}{
\begin{tabular}{lll}
\hline\hline
Quantity & Value & Comment \\
\hline
Reference volume $\Omega$ & 63.293954~\AA$^3$ & Cubic PBEsol reference \\
Electronic dielectric tensor $\epsilon_\infty$ & $6.869406\,I$ & Cubic electronic screening \\
Mode effective charge $Z$ & $19.76637\,I\,e$ & Cubic soft-mode projection \\
Polarization derivative $\partial P/\partial u$ & $5.00351\,I$~C~m$^{-2}$~\AA$^{-1}$ & $Z/\Omega$ \\
Projected well amplitude $u_0$ & 0.07158677~\AA & Projection of fixed-cell tetragonal endpoint \\
Projection residual & 0.4455 & Relative residual of full tetragonal endpoint after one-mode projection \\
Berry-phase polarization & 0.346588~C~m$^{-2}$ & Fixed-cell tetragonal endpoint \\
Model polarization & 0.3582~C~m$^{-2}$ & $(\partial P/\partial u)u_0$ \\
Polarization mismatch & 3.35\% & Model estimate relative to Berry-phase endpoint \\
Double-well barrier $\Delta E_b$ & 38.43446~meV/f.u. & Fixed-cell tetragonal switching path \\
Comparable PBEsol barrier & 34.63~meV & Ref.~\cite{Esswein2022PRResearch_BTO}; different coordinate and reference \\
Comparable PBE barrier & 104.8~meV & Ref.~\cite{Esswein2022PRResearch_BTO}; different functional, coordinate, and reference \\
$A_2$ & $-29.99958791$~eV~\AA$^{-2}$ & Stable quartic, exactly barrier matched \\
$B_4$ & $5.85396071\times10^3$~eV~\AA$^{-4}$ & Stable quartic, exactly barrier matched \\
$C_6$ & 0 & Stable quartic model \\
Barrier-matched quartic RMSE & 0.76972~meV & Error against DFT path points \\
$\kappa_\perp$ & 0.1~eV~\AA$^{-2}$ & Weak transverse simulation parameter \\
\hline\hline
\end{tabular}}
\end{table*}

The projection residual in Table~\ref{tab:bto_local_model_validation} is defined as
\begin{equation}
r_\mathrm{proj}=\frac{\|\Delta\mathbf r-u_0\mathbf v\|}{\|\Delta\mathbf r\|},
\end{equation}
using the same Cartesian displacement convention as the soft-mode projection. Its nonzero value means that the fixed-cell tetragonal endpoint is not exactly reconstructed by one cubic soft-mode coordinate; it contains secondary internal-strain and strain-coupled components. The cubic coordinate should therefore be viewed as a reduced local-mode representation of the dominant polar distortion, not as a full reconstruction of the relaxed tetragonal structure. This is the same modeling philosophy used in effective-Hamiltonian descriptions: the low-energy polar coordinate carries the polarization, kinetic energy, angular momentum, and long-range dipole coupling, while the remaining microscopic degrees of freedom are integrated into effective onsite and intercell interactions.

The short-range intercell kernel $K_\mathrm{sr}(\mathbf R)$ is obtained by projecting the cubic $3\times3\times3$ interatomic force constants onto the cubic soft-mode basis. In the exported simulation parameter set, the kernel used with explicit dipoles is the zero-sum short-range residual in pair-difference form,
\begin{equation}
\mathbf F^\mathrm{sr}_i=-\sum_{\mathbf R}K_\mathrm{sr}(\mathbf R)\left(\mathbf u_{i+\mathbf R}-\mathbf u_i\right),
\qquad
\sum_{\mathbf R}K_\mathrm{sr}(\mathbf R)\simeq0 .
\end{equation}
The explicit long-range dipolar force is evaluated separately in reciprocal space as
\begin{equation}
K_\mathrm{dip}(\mathbf q)=\frac{4\pi k_e}{\Omega}\frac{\left(Z^{T}\mathbf q\right)\left(Z^{T}\mathbf q\right)^T}{\mathbf q^T\epsilon_\infty\mathbf q},\qquad K_\mathrm{dip}(\mathbf q=0)=0 .
\end{equation}
This separation avoids double counting the long-range Coulomb contribution when the FFT dipolar interaction is included in the nonequilibrium dynamics. The resulting simulation model is therefore a cubic-reference local-mode model whose onsite anharmonic potential is calibrated from a fixed-cell tetragonal DFT double-well path. It preserves the high-symmetry soft-mode basis and cubic dipolar response while matching the DFT ferroelectric well depth, equilibrium local-mode amplitude, and Berry-phase polarization scale of the tetragonal ferroelectric state.

\subsection{Simulation geometry and observables}

The nonequilibrium molecular-dynamics simulations are carried out on a lattice of \(N_x\times N_y\times N_z = 101\times85\times13\) primitive cells. Periodic boundary conditions are imposed along the ferroelectric axis \(x\) and along \(z\), while the transverse direction \(y\) is kept finite. This slab geometry permits lateral redistribution across the finite direction, so that polarization accumulation at the sample edges can be resolved directly.

The thermal drive is imposed through a band-like Langevin temperature profile along \(x\). A central hot region is coupled to a bath at \(T_\mathrm{hot}=50~\mathrm{K}\), while the left and right cold regions are kept near zero temperature, \(T_\mathrm{cold}=10^{-4}~\mathrm{K}\). The hot band has a fractional width of 0.10 of the simulation length along \(x\), with a smoothing fraction of 0.05 at the band edges. This profile produces two oppositely directed longitudinal temperature gradients on the two sides of the hot region, allowing odd-in-gradient transverse responses to be identified from the corresponding opposite edge accumulations.

The magnetic-field-induced term is included through the effective gyroscopic coupling defined above, with \(\mathbf{B}\parallel\hat{z}\). Its strength is specified by the phenomenological rate \(\Omega_B=80~\mathrm{THz}\). Simulations are performed for zero, positive, and negative gyroscopic coupling, and the field-odd response is extracted as \(\Delta_B X=[X(+\Omega_B)-X(-\Omega_B)]/2\). The reported nonequilibrium maps are averaged over 200 independent realizations with different random seeds.

The bulk Langevin damping is \(\gamma_\mathrm{bulk}=2.0~\mathrm{ps}^{-1}\). Near the open transverse boundaries, a 10-layer sponge region is applied along \(y\), where the damping is increased toward \(\gamma_\mathrm{edge}=5.0~\mathrm{ps}^{-1}\) to improve stability and suppress drift. The nominal integration time step is \(\Delta t=0.005~\mathrm{ps}\). Each time step is subdivided into two equal substeps of duration \(0.0025~\mathrm{ps}\), and the symmetric conservative, gyroscopic, and Langevin updates are applied within each substep. The gyroscopic part is treated as an exact velocity-space rotation, while the stochastic Langevin part is propagated with the exact Ornstein--Uhlenbeck update. Each trajectory is evolved for \(t_\mathrm{max}=2000~\mathrm{ps}\), with the first \(t_\mathrm{equil}=300~\mathrm{ps}\) discarded as equilibration. Observables are sampled every 10 integration steps, corresponding to a sampling interval of \(0.05~\mathrm{ps}\), and accumulated into blocks of 1000 samples for statistical averaging.

The spatially resolved observables are computed directly from the projected local-mode dynamics. The local polarization follows from
\[
\vect{P}_i=\frac{1}{\Omega}\matr{Z}\vect{u}_i,
\]
the local kinetic energy from
\[
E_{\mathrm{kin},i}=\frac{1}{2}\dot{\vect{u}}_i^{\,T}\matr{M}\dot{\vect{u}}_i,
\]
and the local phonon angular momentum from
\[
L_{i,k}=\vect{u}_i^T\matr{\Lambda}^{(k)}\dot{\vect{u}}_i.
\]
At zero magnetic field, the thermal profile produces a redistribution of kinetic energy, a suppression of the ferroelectric order parameter in the hotter region, and a transverse texture of phonon angular momentum. When the magnetic term is applied for an out-of-plane field, the nonequilibrium steady state develops an additional field-odd redistribution in the energy and polarization channels, giving the phonon Hall-type and ferron Hall responses discussed in the main text.

\begin{figure*}[t]
    \centering
    \includegraphics[width=0.95\textwidth]{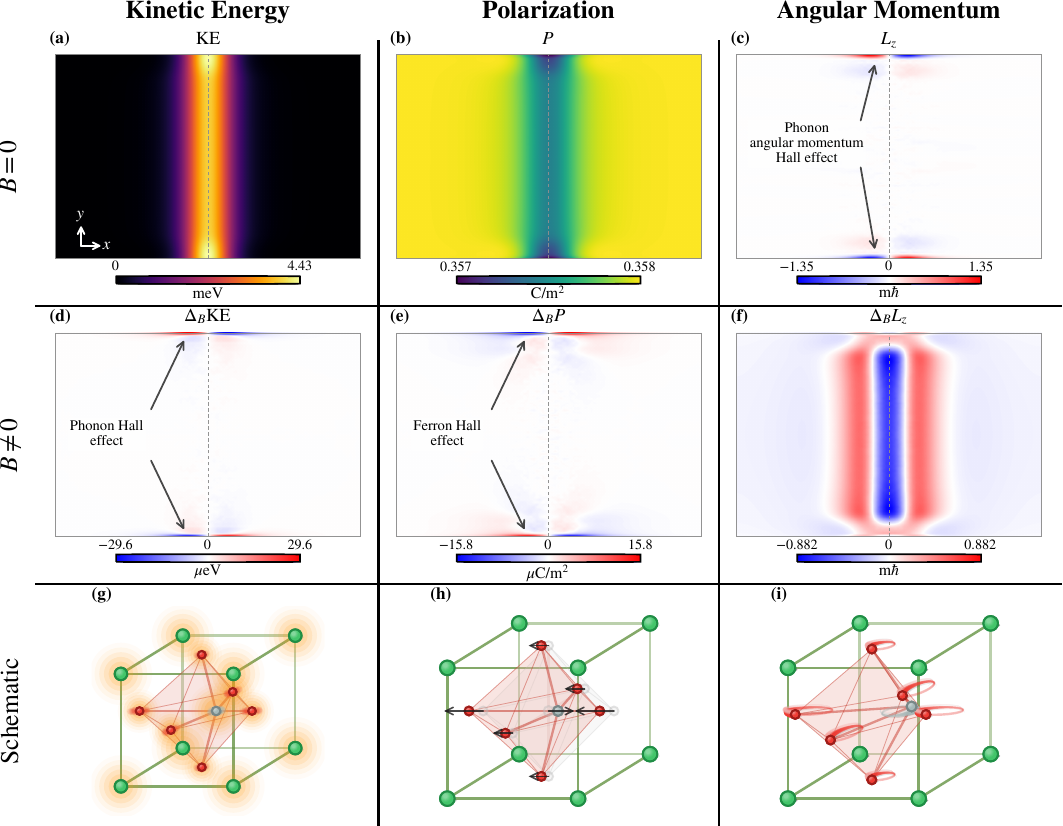}
    \caption{
    \textbf{Channel-resolved nonequilibrium response of the BaTiO$_3$ local-mode simulation.}
    \textbf{(a--c)} Zero-field steady-state maps of the local kinetic energy KE, longitudinal polarization \(P\), and out-of-plane phonon angular momentum \(L_z\).
    \textbf{(d--f)} Field-odd response maps for \(\mathbf B\parallel\hat z\).
    \textbf{(g--i)} Microscopic illustrations of the kinetic-energy, polarization, and angular-momentum channels of the polar soft mode. Green, red, and gray spheres denote Ba, O, and Ti atoms, respectively, and the translucent polyhedron represents the TiO$_6$ octahedron.
    }
    \label{fig:S_BTO_profiles}
\end{figure*}

Figure~\ref{fig:S_BTO_profiles} shows the full channel-resolved nonequilibrium response of the BaTiO$_3$ local-mode simulation. We plot the local kinetic energy, the longitudinal polarization \(P=\hat{\mathbf e}_{\parallel}\cdot\mathbf P\), and the out-of-plane phonon angular momentum \(L_z\). These maps show how the same driven soft-mode dynamics appears in the energy, polarization, and angular-momentum channels.

Figures~\ref{fig:S_BTO_profiles}(a)--\ref{fig:S_BTO_profiles}(c) show the steady state without magnetic field. The imposed temperature profile produces a hot central band, visible as an enhanced kinetic energy in Fig.~\ref{fig:S_BTO_profiles}(a). In the same region, the polarization \(P\) is reduced because stronger soft-mode fluctuations suppress the average ferroelectric distortion. The angular-momentum map in Fig.~\ref{fig:S_BTO_profiles}(c) shows opposite signs of \(L_z\) accumulating near opposite transverse \(y\) edges. The sign reverses on the two sides of the hot band, where the longitudinal temperature gradient changes direction. This is the phonon-angular-momentum Hall response: a temperature-gradient-driven transverse accumulation of local circular or elliptical lattice motion.

We then apply the gyroscopic magnetic coupling with \(\mathbf B\parallel\hat z\) and isolate the component that is odd under field reversal, $\Delta_B X$. The resulting maps are shown in Figs.~\ref{fig:S_BTO_profiles}(d)--\ref{fig:S_BTO_profiles}(f). The field-odd kinetic-energy response \(\Delta_B\mathrm{KE}\), Fig.~\ref{fig:S_BTO_profiles}(d), shows opposite-sign accumulation near the transverse edges. This is the phonon-Hall-type response in the energy channel: once time-reversal symmetry is broken by the magnetic coupling, the thermally driven soft-mode motion is deflected sideways and redistributes vibrational energy.

The field-odd polarization response in Fig.~\ref{fig:S_BTO_profiles}(e) is the ferron Hall signal. Since the driven soft mode carries electric polarization, the same magnetic deflection also redistributes the polarization-carrying coordinate. The resulting \(\Delta_B P\) accumulates with opposite signs at the transverse edges and reverses between the two sides of the hot band, following the reversal of the longitudinal temperature gradient. This odd-in-field and odd-in-gradient transverse accumulation of the ferroelectric order parameter is the ferron Hall effect.

Figure~\ref{fig:S_BTO_profiles}(f) shows the corresponding field-odd change in \(L_z\). The zero-field angular-momentum texture is generated by the thermal drive, while the magnetic coupling produces an additional field-odd redistribution. The three lower panels summarize the microscopic meaning of the channels: kinetic-energy modulation corresponds to enhanced soft-mode fluctuations, polarization response to a shift of the soft-mode coordinate along the ferroelectric direction, and angular momentum to circular or elliptical ionic motion carrying \(L_z\). Thus Fig.~\ref{fig:S_BTO_profiles} separates the phonon-angular-momentum Hall response, the phonon-Hall-type energy response, and the ferron Hall polarization response within the same driven polar soft-mode dynamics.

\subsection{Soft-mode energy transport, angular momentum transport, and ferron Hall polarization shift}
\label{sec:softmode_transport_balances}

Starting from the local-mode equation of motion, we derive the balances that connect the nonequilibrium maps to transport. The kinetic-energy balance identifies how the thermal baths and the conservative forces change the local velocity fluctuations. Adding the conservative potential-energy terms then gives the intercell soft-mode energy current. Applying the same procedure to the projected angular momentum gives an angular-momentum current and the associated local torques. Finally, projecting the steady force balance along the ferroelectric axis gives the field-odd polarization shift inside the anharmonic well.

The projected soft-mode dynamics is governed by
\begin{equation}
\matr{M}\ddot{\vect{u}}_i
=
\vect{F}^{\mathrm{on}}_i
+
\vect{F}^{\mathrm{sr}}_i
+
\vect{F}^{\mathrm{dip}}_i
+
\matr{G}_{\Omega_B}\dot{\vect{u}}_i
-
\gamma_i\matr{M}\dot{\vect{u}}_i
+
\vect{\eta}_i(t).
\label{eq:transport_softmode_eom}
\end{equation}
The local kinetic energy is
\begin{equation}
E_{\mathrm{kin},i}
=
\frac{1}{2}
\dot{\vect{u}}_i^T
\matr{M}
\dot{\vect{u}}_i .
\label{eq:transport_ekin_def}
\end{equation}
Taking its time derivative and using Eq.~\eqref{eq:transport_softmode_eom} gives
\begin{align}
\frac{dE_{\mathrm{kin},i}}{dt}
=&\,
\dot{\vect{u}}_i^T
\vect{F}^{\mathrm{on}}_i
+
\dot{\vect{u}}_i^T
\vect{F}^{\mathrm{sr}}_i
+
\dot{\vect{u}}_i^T
\vect{F}^{\mathrm{dip}}_i
+
\dot{\vect{u}}_i^T
\matr{G}_{\Omega_B}
\dot{\vect{u}}_i
\nonumber\\
&-
\gamma_i
\dot{\vect{u}}_i^T
\matr{M}
\dot{\vect{u}}_i
+
\dot{\vect{u}}_i^T
\vect{\eta}_i(t).
\label{eq:transport_ekin_balance_raw}
\end{align}
The gyroscopic matrix is antisymmetric, so
\begin{equation}
\dot{\vect{u}}_i^T
\matr{G}_{\Omega_B}
\dot{\vect{u}}_i
=
0 .
\label{eq:transport_gyro_no_work}
\end{equation}
The magnetic coupling therefore rotates the soft-mode velocity without doing work. It can deflect the energy flow generated by the thermal drive, but it does not inject kinetic energy locally.

The short-range force is resolved into pair forces as
\begin{equation}
\vect{F}^{\mathrm{sr}}_i
=
\sum_{\vect{R}}
\vect{F}^{\mathrm{sr}}_{i\leftarrow i+\vect{R}},
\qquad
\vect{F}^{\mathrm{sr}}_{i\leftarrow i+\vect{R}}
=
-
\matr{K}_{\mathrm{sr}}(\vect{R})
\left(
\vect{u}_{i+\vect{R}}-\vect{u}_i
\right).
\label{eq:transport_pair_force}
\end{equation}
The kinetic-energy change due to the short-range couplings is then the power delivered by these pair forces,
\begin{equation}
P^{\mathrm{sr}}_{\mathrm{kin},i}
=
\sum_{\vect{R}}
\left\langle
\dot{\vect{u}}_i^T
\vect{F}^{\mathrm{sr}}_{i\leftarrow i+\vect{R}}
\right\rangle .
\label{eq:transport_kinetic_power_sr}
\end{equation}
The onsite and dipolar powers are
\begin{align}
P^{\mathrm{on}}_{\mathrm{kin},i}
&=
\left\langle
\dot{\vect{u}}_i^T
\vect{F}^{\mathrm{on}}_i
\right\rangle ,
\label{eq:transport_kinetic_power_on}
\\
P^{\mathrm{dip}}_{\mathrm{kin},i}
&=
\left\langle
\dot{\vect{u}}_i^T
\vect{F}^{\mathrm{dip}}_i
\right\rangle .
\label{eq:transport_kinetic_power_dip}
\end{align}
Averaging over the Langevin noise gives
\begin{equation}
\frac{d\langle E_{\mathrm{kin},i}\rangle}{dt}
=
P^{\mathrm{on}}_{\mathrm{kin},i}
+
P^{\mathrm{sr}}_{\mathrm{kin},i}
+
P^{\mathrm{dip}}_{\mathrm{kin},i}
-
\gamma_i
\left\langle
\dot{\vect{u}}_i^T
\matr{M}
\dot{\vect{u}}_i
\right\rangle
+
3\gamma_i k_B T_i .
\label{eq:transport_kinetic_balance}
\end{equation}
The conservative forces exchange energy between the kinetic and potential sectors. The damping term removes kinetic energy, while the stochastic term injects kinetic energy into the three soft-mode components. In the steady state, the left-hand side of Eq.~\eqref{eq:transport_kinetic_balance} vanishes. The field-odd kinetic-energy map is therefore a map of how the gyroscopic deflection redistributes the velocity fluctuations sustained by the thermal reservoirs.

A transport current is obtained after including the conservative potential energy. Assigning one half of each short-range bond energy to each endpoint cell gives a local soft-mode energy $\varepsilon_i$ containing the kinetic energy, the onsite energy, and the assigned short-range intercell energy. The short-range energy current from cell $i$ to cell $i+\vect{R}$ is
\begin{equation}
J^E_{i\rightarrow i+\vect{R}}
=
-
\frac{1}{2}
\left\langle
\left(
\dot{\vect{u}}_i+\dot{\vect{u}}_{i+\vect{R}}
\right)^T
\vect{F}^{\mathrm{sr}}_{i\leftarrow i+\vect{R}}
\right\rangle .
\label{eq:transport_energy_current}
\end{equation}
With this definition, the local soft-mode energy balance becomes
\begin{equation}
\frac{d\langle\varepsilon_i\rangle}{dt}
+
\sum_{\vect{R}}
J^E_{i\rightarrow i+\vect{R}}
=
\left\langle
\dot{\vect{u}}_i^T
\vect{F}^{\mathrm{dip}}_i
\right\rangle
+
P^{\mathrm{bath}}_i ,
\label{eq:transport_energy_balance}
\end{equation}
where
\begin{equation}
P^{\mathrm{bath}}_i
=
-
\gamma_i
\left\langle
\dot{\vect{u}}_i^T
\matr{M}
\dot{\vect{u}}_i
\right\rangle
+
3\gamma_i k_B T_i .
\label{eq:transport_bath_power}
\end{equation}
The dipolar force is retained as a nonlocal power term because the dipolar interaction is evaluated in reciprocal space. Equation~\eqref{eq:transport_energy_current} is the short-range contribution to the soft-mode energy current. Its field-odd transverse component is the energy-current response associated with the phonon Hall deflection of the polar soft mode.

The projected angular momentum of cell $i$ is
\begin{equation}
L_{i,k}
=
\vect{u}_i^T
\matr{\Lambda}^{(k)}
\dot{\vect{u}}_i ,
\qquad
k\in\{x,y,z\}.
\label{eq:transport_L_def}
\end{equation}
Taking the derivative gives
\begin{equation}
\frac{dL_{i,k}}{dt}
=
\dot{\vect{u}}_i^T
\matr{\Lambda}^{(k)}
\dot{\vect{u}}_i
+
\vect{u}_i^T
\matr{\Lambda}^{(k)}
\ddot{\vect{u}}_i .
\label{eq:transport_L_derivative}
\end{equation}
Since $\matr{\Lambda}^{(k)}$ is antisymmetric,
\begin{equation}
\dot{\vect{u}}_i^T
\matr{\Lambda}^{(k)}
\dot{\vect{u}}_i
=
0.
\label{eq:transport_lambda_antisym}
\end{equation}
Using Eq.~\eqref{eq:transport_softmode_eom}, the angular-momentum balance becomes
\begin{align}
\frac{dL_{i,k}}{dt}
=&\,
\vect{u}_i^T
\matr{\Lambda}^{(k)}
\matr{M}^{-1}
\vect{F}^{\mathrm{on}}_i
+
\vect{u}_i^T
\matr{\Lambda}^{(k)}
\matr{M}^{-1}
\vect{F}^{\mathrm{sr}}_i
+
\vect{u}_i^T
\matr{\Lambda}^{(k)}
\matr{M}^{-1}
\vect{F}^{\mathrm{dip}}_i
\nonumber\\
&+
\vect{u}_i^T
\matr{\Lambda}^{(k)}
\matr{M}^{-1}
\matr{G}_{\Omega_B}
\dot{\vect{u}}_i
-
\gamma_i L_{i,k}
+
\vect{u}_i^T
\matr{\Lambda}^{(k)}
\matr{M}^{-1}
\vect{\eta}_i(t).
\label{eq:transport_L_balance_raw}
\end{align}
The short-range force transfers angular momentum between neighboring local modes. We define the corresponding bond current by
\begin{equation}
J^{L_k}_{i\rightarrow i+\vect{R}}
=
-
\left\langle
\vect{u}_i^T
\matr{\Lambda}^{(k)}
\matr{M}^{-1}
\vect{F}^{\mathrm{sr}}_{i\leftarrow i+\vect{R}}
\right\rangle .
\label{eq:transport_L_current}
\end{equation}
After averaging over the stochastic force, whose mean vanishes, one obtains
\begin{equation}
\frac{d\langle L_{i,k}\rangle}{dt}
+
\sum_{\vect{R}}
J^{L_k}_{i\rightarrow i+\vect{R}}
=
\tau^{\mathrm{on}}_{i,k}
+
\tau^{\mathrm{dip}}_{i,k}
+
\tau^{B}_{i,k}
-
\gamma_i\langle L_{i,k}\rangle .
\label{eq:transport_L_balance}
\end{equation}
The local torques are
\begin{align}
\tau^{\mathrm{on}}_{i,k}
&=
\left\langle
\vect{u}_i^T
\matr{\Lambda}^{(k)}
\matr{M}^{-1}
\vect{F}^{\mathrm{on}}_i
\right\rangle ,
\label{eq:transport_tau_on}
\\
\tau^{\mathrm{dip}}_{i,k}
&=
\left\langle
\vect{u}_i^T
\matr{\Lambda}^{(k)}
\matr{M}^{-1}
\vect{F}^{\mathrm{dip}}_i
\right\rangle ,
\label{eq:transport_tau_dip}
\\
\tau^{B}_{i,k}
&=
\left\langle
\vect{u}_i^T
\matr{\Lambda}^{(k)}
\matr{M}^{-1}
\matr{G}_{\Omega_B}
\dot{\vect{u}}_i
\right\rangle .
\label{eq:transport_tau_B}
\end{align}
The angular momentum is therefore transported by the intercell short-range force, damped by the Langevin bath, and modified locally by onsite, dipolar, and magnetic torques.

The polarization is obtained from the same coordinate,
\begin{equation}
\vect{P}_i
=
\frac{1}{\Omega}
\matr{Z}
\vect{u}_i .
\label{eq:transport_P_def}
\end{equation}
Its time derivative is
\begin{equation}
\dot{\vect{P}}_i
=
\frac{1}{\Omega}
\matr{Z}
\dot{\vect{u}}_i .
\label{eq:transport_Pdot}
\end{equation}
In the nonequilibrium steady state, $\langle\dot{\vect{P}}_i\rangle=0$. The static ferron Hall signal is therefore the field-odd shift of the local polar coordinate,
\begin{equation}
\Delta_B\vect{P}_i
=
\frac{1}{2}
\left[
\vect{P}_i(+\Omega_B)
-
\vect{P}_i(-\Omega_B)
\right]
=
\frac{1}{\Omega}
\matr{Z}
\Delta_B
\langle\vect{u}_i\rangle .
\label{eq:transport_deltaP_def}
\end{equation}

The sign of this shift follows from the anharmonic force balance inside one ferroelectric well. For a state polarized along $\hat{\vect{e}}_{\parallel}$, write
\begin{equation}
u_{i,\parallel}
=
\hat{\vect{e}}_{\parallel}\cdot\vect{u}_i
=
u_0+\delta u_{i,\parallel}.
\label{eq:transport_delta_u_def}
\end{equation}
The longitudinal onsite potential is
\begin{equation}
V_{\parallel}(u)
=
\frac{A_2}{2}u^2
+
\frac{B_4}{4}u^4
+
\frac{C_6}{6}u^6 .
\label{eq:transport_longitudinal_potential}
\end{equation}
The minimum satisfies
\begin{equation}
A_2+B_4u_0^2+C_6u_0^4=0.
\label{eq:transport_minimum_condition}
\end{equation}
Expanding about $u_0$ gives
\begin{equation}
V_{\parallel}(u_0+\delta u)
=
V_{\parallel}(u_0)
+
\frac{k_{\parallel}}{2}
(\delta u)^2
+
\frac{a_{\parallel}}{6}
(\delta u)^3
+
\cdots ,
\label{eq:transport_well_expansion}
\end{equation}
with
\begin{align}
k_{\parallel}
&=
A_2+3B_4u_0^2+5C_6u_0^4,
\label{eq:transport_k_parallel}
\\
a_{\parallel}
&=
6B_4u_0+20C_6u_0^3 .
\label{eq:transport_a_parallel}
\end{align}
The longitudinal onsite force near the bottom of the well is
\begin{equation}
F^{\mathrm{on}}_{i,\parallel}
=
-
k_{\parallel}
\delta u_{i,\parallel}
-
\frac{a_{\parallel}}{2}
(\delta u_{i,\parallel})^2
+
\cdots .
\label{eq:transport_longitudinal_force}
\end{equation}
Taking the steady-state average of the longitudinal equation of motion gives
\begin{equation}
0
=
-
k_{\parallel}
\langle\delta u_{i,\parallel}\rangle
-
\frac{a_{\parallel}}{2}
\left\langle
(\delta u_{i,\parallel})^2
\right\rangle
+
\left\langle
F^{\mathrm{sr}}_{i,\parallel}
+
F^{\mathrm{dip}}_{i,\parallel}
\right\rangle
+
\cdots .
\label{eq:transport_mean_force_balance}
\end{equation}
The damping, gyroscopic, and stochastic terms do not contribute to this mean static force balance because $\langle\dot{\vect{u}}_i\rangle=0$ and $\langle\vect{\eta}_i\rangle=0$. If the local anharmonic force gives the leading static shift, Eq.~\eqref{eq:transport_mean_force_balance} reduces to
\begin{equation}
\langle\delta u_{i,\parallel}\rangle
\simeq
-
\frac{a_{\parallel}}{2k_{\parallel}}
\left\langle
(\delta u_{i,\parallel})^2
\right\rangle .
\label{eq:transport_mean_shift_variance}
\end{equation}
The cubic anharmonicity of the well converts fluctuation intensity into a mean displacement of the polar coordinate. In the positive ferroelectric well, increasing the longitudinal fluctuation amplitude shifts the average coordinate toward the paraelectric saddle and reduces the polarization.

The longitudinal polarization is
\begin{equation}
P_{i,\parallel}
=
P_0
+
\frac{Z_{\parallel}}{\Omega}
\delta u_{i,\parallel},
\qquad
Z_{\parallel}
=
\hat{\vect{e}}_{\parallel}^{T}
\matr{Z}
\hat{\vect{e}}_{\parallel}.
\label{eq:transport_longitudinal_P}
\end{equation}
Combining Eqs.~\eqref{eq:transport_mean_shift_variance} and \eqref{eq:transport_longitudinal_P} gives
\begin{equation}
\Delta_B P_{i,\parallel}
\simeq
-
\frac{Z_{\parallel}}{\Omega}
\frac{a_{\parallel}}{2k_{\parallel}}
\Delta_B
\left\langle
(\delta u_{i,\parallel})^2
\right\rangle .
\label{eq:transport_deltaP_variance}
\end{equation}
The field-odd polarization accumulation has the opposite sign to the field-odd redistribution of the longitudinal fluctuation variance.

The same relation can be expressed in terms of the longitudinal kinetic energy,
\begin{equation}
E^{\parallel}_{\mathrm{kin},i}
=
\frac{1}{2}
M_{\parallel}
\left\langle
\dot{u}_{i,\parallel}^2
\right\rangle ,
\qquad
M_{\parallel}
=
\hat{\vect{e}}_{\parallel}^{T}
\matr{M}
\hat{\vect{e}}_{\parallel}.
\label{eq:transport_longitudinal_ekin}
\end{equation}
For fluctuations governed by the harmonic part of the well,
\begin{equation}
E^{\parallel}_{\mathrm{kin},i}
\simeq
\frac{1}{2}
k_{\parallel}
\left\langle
(\delta u_{i,\parallel})^2
\right\rangle .
\label{eq:transport_equipartition}
\end{equation}
Substituting Eq.~\eqref{eq:transport_equipartition} into Eq.~\eqref{eq:transport_deltaP_variance} gives
\begin{equation}
\Delta_B P_{i,\parallel}
\simeq
-
\frac{Z_{\parallel}}{\Omega}
\frac{a_{\parallel}}{k_{\parallel}^2}
\Delta_B E^{\parallel}_{\mathrm{kin},i}.
\label{eq:transport_deltaP_ekin}
\end{equation}
Regions where the magnetic field increases the longitudinal soft-mode kinetic energy therefore show a reduced polarization, while regions where the field depletes the longitudinal kinetic energy show an enhanced polarization. The ferron Hall signal is the polarization response of the anharmonic ferroelectric well to the phonon-Hall-like redistribution of kinetic energy in the polar soft-mode sector.

\clearrevtexfrontmatternotes
\putbib[refs]
\clearrevtexfrontmatternotes
\end{bibunit}

\end{document}